\documentstyle[12pt,preprint,eqsecnum,aps,epsfig]{revtex}
\def\nn{\nonumber}
\def\dslash{\hspace{-0.2cm} \slash \hspace{0.1cm}}
\def\ra{\langle r^2 \rangle_A}
\tighten
\begin{document}
\preprint{MKPH-T-03-4}
\draft
\title{Pion electroproduction, PCAC, chiral Ward identities, 
and the axial form factor revisited}
\author{T. Fuchs and S. Scherer}
\address{Institut f\"ur Kernphysik, Johannes Gutenberg-Universit\"at,
D-55099 Mainz, Germany}
\date{March 1, 2003}
\maketitle
\begin{abstract}
   We re-investigate Adler's PCAC relation in the presence of an external 
electromagnetic field within the framework of QCD coupled to external fields.
   We discuss pion electroproduction within a tree-level approximation
to chiral perturbation theory and explicitly verify a chiral Ward identity
referred to as the Adler-Gilman relation.
   We critically examine soft-momentum techniques and point out how 
inadmissable approximations may lead to results incompatible with
chiral symmetry. 
   As a result we confirm that threshold pion electroproduction is indeed
a tool to obtain information on the axial form factor of the nucleon.
\end{abstract}

\section{Introduction}
\label{section_introduction}
    Pion photo- and electroproduction both have a long tradition as a tool to
obtain information on strong as well as electroweak properties
of the pion and the nucleon.
   For example, as early as 1954 Kroll and Ruderman \cite{Kroll:1954}, 
in their famous low-energy theorem (LET), discussed the possibility of 
extracting the renormalized pion-nucleon coupling constant $g_{\pi N}$ 
from charged pion photoproduction at threshold.
   Besides Lorentz covariance, the essential ingredient entering the 
derivation of the Kroll-Ruderman theorem was the application of the Ward 
identity \cite{Ward:1950xp} resulting from gauge invariance.

   The algebra of currents and the hypothesis of a partially conserved 
axial-vector current (PCAC) resulted in additional constraints such as
the theorem of Fubini, Furlan, and Rossetti \cite{Fubini:1965} establishing
a sum rule for pion photoproduction in terms of the anomalous isovector and 
isoscalar magnetic moments of the nucleon.
   The potential of investigating the axial-vector form factor of the 
nucleon through the electroproduction of charged pions was first realized
by Nambu and Shrauner in the framework of the chirality formalism 
\cite{Nambu:1962wa,Nambu:1962wb}.
   Subsequently, their result has been recovered and extended within various
approaches 
\cite{Adler:1966,Gleeson:zp,Vainshtein:ih,Dombey:1973ve,Scherer:1991cy}
(for an overview, see Ref.\ \cite{Amaldi:1979vh}).

   The current-algebra and PCAC approaches of the 1960's (see, e.g.,
Refs.\ \cite{Adler:1968,Treiman:1972,Alfaro:1973})
had in common that they made no explicit reference to the dynamical origin 
of the underlying symmetry.
   In our present understanding, the symmetry currents originate in a
global chiral $\mbox{SU}(N_f)_L\times \mbox{SU}(N_f)_R$ invariance 
of QCD for $N_f$ massless quark flavors.
   As already pointed out by Gell-Mann \cite{Gell-Mann:1962xb}, 
even if a continuous symmetry is violated by large effects, it will still have 
some physical consequences, which can be studied if the 
symmetry breaking pattern is explicitly known.
   In the present context, the symmetry breaking in question is associated with
the finite quark masses.
   It is rather straightforward to derive the so-called chiral Ward identities 
among QCD Green functions implied by the symmetry currents and the
symmetry-breaking pattern, while it is more difficult to actually satisfy 
these constraints in practical calculations.
   However, in the framework of effective field theory, the chiral Ward 
identities will automatically be satisfied if the underlying chiral symmetry 
(and its breaking pattern) 
is systematically mapped onto the most general effective Lagrangian in terms 
of the relevant experimentally observed degrees of freedom
\cite{Weinberg:1979kz,Gasser:1984yg,Gasser:1984gg,Gasser:1988rb}.
   Turning this mapping into useful consequences requires a method
which allows for a rigorous analysis of a particular contribution to 
a Green function in terms of some expansion scheme.
   This is provided by Weinberg's power counting 
\cite{Weinberg:1979kz,Weinberg:1991um}
which makes use of the special role played by the pion as
the approximate Goldstone boson of spontaneous chiral symmetry breaking.
   Its weak coupling to other hadrons in the low-energy limit, in combination
with its small mass, allow for an analysis of the low-energy structure of
QCD Green functions in the framework of chiral perturbation
theory (ChPT) 
\cite{Weinberg:1979kz,Gasser:1984yg,Gasser:1984gg,Gasser:1988rb}.
   In the single-nucleon sector a consistent power counting has been
developed for both the so-called heavy-baryon formulation 
\cite{Jenkins:1990jv,Bernard:1992qa} and, more recently, also for the 
relativistic approach  \cite{Ellis:1997kc,Becher:1999he,Fuchs:2003qc}.

   The present work aims at shedding additional light on the importance of
chiral Ward identities in the context of extracting the axial form factor
of the nucleon from pion electroproduction experiments
\cite{Choi:1993vt,Blomqvist:tx,Liesenfeld:1999mv,Bartsch:1998,Mueller:2002,%
Baumann:2003}.
   We will first show that, within the framework of QCD coupled to external 
fields, the PCAC relation for a particular choice
of the pion interpolating field is of the same form as the one originally
obtained by Adler \cite{Adler:1965} through minimal substitution.
   This is due to the fact that, within QCD, the quark fields entering
the symmetry currents are fundamental, i.e., point-like degrees of freedom.

   Our subsequent discussion of the chiral Ward identities will be performed 
in the framework of a tree-level approximation to chiral perturbation theory 
in order to keep the line of arguments as transparent as possible.
   In terms of a loop expansion such tree-level diagrams may be understood
as the leading order in an expansion in terms of $\hbar$ 
\cite{Ryder:wq,Zinn-Justin:1989mi}.
   Moreover, chiral Ward identities are expected to be satisfied order
by order in the loop expansion \cite{Becher:1999he,Fuchs:2003qc}.
   Of course, ChPT also allows one to systematically evaluate corrections to
the tree-level results.
   In the context of pion photo- and electroproduction this was done
in a series of papers by Bernard {\em et al.} 
\cite{Bernard:1991rt,Bernard:1992ys,Bernard:1993rf,Bernard:2001rs}.
   
   In the effective-field-theory approach we will point out
the distinction between the chiral Ward identities and the so-called 
electromagnetic Ward-Takahashi identities 
\cite{Ward:1950xp,Fradkin:1955jr,Takahashi:xn} 
applying to the effective degrees of freedom.
   The origin of these additional identities resides in the fact that the
effective hadronic degrees of freedom, namely pions and nucleons, are
carriers of U(1) representations.
   As a consequence, the building blocks of a calculation in the effective
theory also have to satisfy these identities.
 
   Our approach will allow us to clarify a discussion triggered by
a paper of Haberzettl \cite{Haberzettl:2000sm}, where it was argued that
in the case of pion electroproduction PCAC does not provide any additional 
constraints beyond the Goldberger-Treiman relation.  
   We will explicitly point out which step in the discussion of
Ref.\ \cite{Haberzettl:2000sm} is problematic
(see also Refs.\ 
\cite{Guichon:2001ew,Bernard:2001ii,Haberzettl:2001yt,Truhlik:2001nn}
for additional discussion).

   Our paper is organized as follows.
   In Sec.\ \ref{section_PCAC} we first re-derive Adler's PCAC relation in 
the framework of the QCD Lagrangian coupled to external fields.
   We then define the relevant Green functions and establish a
chiral Ward identity (Adler-Gilman relation \cite{Adler:1966})
entering pion photo- and electroproduction.
   In Sec.\ \ref{section_effective_lagrangian} we introduce the relevant parts 
of the effective Lagrangians used.
   In Sec.\ \ref{section_axial_vector-current_matrix_elements}
we discuss various matrix elements involving the axial-vector
current and the pseudoscalar density in order to illustrate the simplest
chiral Ward identity, namely the PCAC relation, at work.
   Those readers familiar with the concepts of chiral perturbation
theory are invited to immediately move forward to Sec.\ 
\ref{section_pion_electroproduction} containing 
the central piece of this work, namely, a discussion of the Adler-Gilman 
relation and its relation to the pion production amplitude. 
   We will study the traditional soft-momentum limit and comment on the
method of Ref.\ \cite{Haberzettl:2000sm}.
   Finally, we will explain how, in terms of the effective degrees
of freedom, the constraints due to the chiral symmetry of QCD 
are translated into relations among the vertices of the effective theory. 
   General conclusions are presented in Sec.\ \ref{section_conclusions}.

\section{The PCAC relation in the presence of external fields}
\label{section_PCAC}
   For $N_f$ massless quarks, the QCD Hamiltonian is invariant under the 
operation of the chiral group $\mbox{SU}(N_f)_L\times \mbox{SU}(N_f)_R$ on 
the left- and right-handed quark fields \cite{Gasser:1984yg}.
   Associated with this invariance are $2 (N_f^2-1)$ symmetry currents. 
   Here we will restrict ourselves to a discussion of the approximate 
$\mbox{SU}(2)_L\times \mbox{SU}(2)_R$ symmetry in the sector of 
two light flavors.
   The finite $u$- and $d$-quark masses result in explicit divergences
of the symmetry currents.
   However, as first pointed out by Gell-Mann, the equal-time commutation
relations still play an important role even if the symmetry is explicitly 
broken \cite{Gell-Mann:1962xb}.
   In general, the symmetry currents will lead to chiral Ward identities
relating various QCD Green functions with each other.    

\subsection{The PCAC relation in the presence of external fields}
   The starting point of our discussion is a derivation of the SU(2) PCAC 
relation from the QCD Lagrangian in the presence of general external fields.
   To that end, we consider the two-flavor QCD Lagrangian coupled to external 
c-number fields $v_\mu (x)$, $v_\mu^{(s)} (x)$, $a_\mu (x)$, $s(x)$, and $p(x)$
\cite{Gasser:1984yg}:
\begin{equation}
\label{lqcds}
{\cal L}={\cal L}^0_{\rm QCD}+{\cal L}_{\rm ext}
={\cal L}^0_{\rm QCD}+\bar{q}\gamma^\mu (v_\mu +\frac{1}{3} v_\mu^{(s)}
+\gamma_5 a_\mu )q
-\bar{q}(s-i\gamma_5 p)q,
\end{equation}
where ${\cal L}^0_{\rm QCD}$ refers to the QCD Lagrangian for massless
$u$ and $d$ quarks.\footnote{The remaining flavors $s$, $\dots$, $t$
appear with their respective mass terms.}
    The external fields are color-neutral, Hermitian $2\times 2$ matrices,
where we parameterize the matrix character, with respect to the (suppressed) 
flavor indices $u$ and $d$ of the quark fields, as \cite{Gasser:1984yg}
\begin{mathletters}
\label{externalfields}
\begin{eqnarray}
\label{vmuext}
v_\mu(x)&=&\frac{1}{2}[r_\mu(x)+l_\mu(x)]
=\sum_{i=1}^3\frac{\tau_i}{2}v^i_\mu(x),\\
\label{sext}
s(x)&=&1_{2\times 2}\,s_0(x)+\sum_{i=1}^3 \tau_i s_i(x),\\
\label{amuext}
a_\mu(x)&=&\frac{1}{2}[r_\mu(x)-l_\mu(x)]
=\sum_{i=1}^3\frac{\tau_i}{2}a^i_\mu(x),\\
\label{pext}
p(x)&=&1_{2\times 2}\,p_0(x)+\sum_{i=1}^3 \tau_i p_i(x).
\end{eqnarray}
\end{mathletters}
    Here, we do not consider a coupling to an external axial-vector singlet 
field, because the corresponding singlet axial-vector current has an anomaly
such that the Green functions involving the axial-vector singlet current
are related to Green functions containing the contraction of the
gluon field-strength tensor with its dual. 
   The ordinary two-flavor QCD Lagrangian is recovered by setting 
$v_\mu=v_\mu^{(s)}=a_\mu=p=0$ and 
$s=\mbox{diag}(m_u,m_d)$ in Eq.\ (\ref{lqcds}).  
   For simplicity, we will disregard the mass difference $m_u-m_d$ and
consider the isospin-symmetric case $m_u=m_d=\hat{m}$.

   The Lagrangian of Eq.\ (\ref{lqcds}) is invariant under {\em local }
$\mbox{SU(2)}_L\times \mbox{SU(2)}_R\times \mbox{U}(1)_V$ 
transformations of the left-handed
and right-handed quark fields,
\begin{mathletters}
\label{qtrans}
\begin{eqnarray}
q_L(x)&\equiv&\frac{1}{2}(1-\gamma_5)q(x)\mapsto 
\exp\left(-i\frac{\theta(x)}{3}\right)V_L(x) q_L(x),\label{qltrans}\\
q_R(x)&\equiv&\frac{1}{2}(1+\gamma_5)q(x)\mapsto 
\exp\left(-i\frac{\theta(x)}{3}\right)V_R(x) q_R(x)\label{qrtrans},
\end{eqnarray}
\end{mathletters}
where
\begin{equation}
\label{vlr}
V_{L/R}(x)=\exp\left(-i\sum_{i=1}^3\frac{\tau_i}{2}\theta^i_{L/R}(x)\right),
\end{equation}
   provided the external fields transform as
\begin{mathletters}
\label{localtrans}
\begin{eqnarray}
r_\mu&\mapsto& V_R r_\mu V_R^{\dagger}
+iV_R\partial_\mu V_R^{\dagger},\label{rtrans}\\
l_\mu&\mapsto& V_L l_\mu V_L^{\dagger}
+iV_L\partial_\mu V_L^{\dagger},
\label{ltrans}\\
v_\mu^{(s)}&\mapsto&v_\mu^{(s)}-\partial_\mu\theta,\label{vstrans}\\
s+ip&\mapsto& V_R(s+ip)V_L^{\dagger},\label{spiptrans}\\
s-ip&\mapsto& V_L(s-ip)V_R^{\dagger}\label{smiptrans}.
\end{eqnarray}
\end{mathletters}
   Applying the method of Gell-Mann and L\'{e}vy \cite{Gell-Mann:1960np}
to identify currents and their divergences by investigating the variation
of the Lagrangian under local infinitesimal transformations, 
\begin{mathletters}
\label{gellmannlevy}
\begin{eqnarray}
J^\mu&=&\frac{\partial\delta {\cal L}}{\partial(\partial_\mu\epsilon)},
\label{jmu}\\
\partial_\mu J^\mu&=&\frac{\partial\delta{\cal L}}{\partial\epsilon},
\label{djmu}
\end{eqnarray}
\end{mathletters}
leads to the following expressions for the vector and axial-vector currents,
\begin{mathletters}
\label{vmuamu}
\begin{eqnarray}
V^\mu_i&=&\bar{q}\gamma^\mu\frac{\tau_i}{2}q,\label{vmu}\\
A^\mu_i&=&\bar{q}\gamma^\mu\gamma_5\frac{\tau_i}{2}q.\label{amu}
\end{eqnarray}
\end{mathletters}
  If the external fields are not simultaneously transformed and one considers
a {\em global} chiral transformation only, the divergences of the currents
read
\begin{mathletters}
\label{divav}
\begin{eqnarray}
\label{divv}
\partial_\mu V^\mu_i&=&i\bar{q}\gamma^\mu\left[\frac{\tau_i}{2},v_\mu\right]q
+i\bar{q}\gamma^\mu\gamma_5\left[\frac{\tau_i}{2},a_\mu\right]q
-i\bar{q}\left[\frac{\tau_i}{2},s\right]q
-\bar{q}\gamma_5\left[\frac{\tau_i}{2},p\right]q,\\
\label{diva}
\partial_\mu A^\mu_i&=&i\bar{q}\gamma^\mu\gamma_5
\left[\frac{\tau_i}{2},v_\mu\right]q
+i\bar{q}\gamma^\mu\left[\frac{\tau_i}{2},a_\mu\right]q
+i\bar{q}\gamma_5\left\{\frac{\tau_i}{2},s\right\}q
+\bar{q}\left\{\frac{\tau_i}{2},p\right\}q.
\end{eqnarray}
\end{mathletters}
   In the present case we intend to consider the QCD Lagrangian for 
a finite light quark mass $\hat{m}$ in combination with a coupling to an 
external electromagnetic field ${\cal A}_\mu$ given by\footnote{We use 
natural units $\hbar=c=1$, $e>0$, $e^2/(4\pi)\approx 1/137$.}
\begin{equation}
\label{emcoupling}
-e{\cal A}_\mu J^\mu = -e {\cal A}_\mu
\left(\frac{2}{3}\bar{u}\gamma^\mu u - \frac{1}{3}\bar{d}\gamma^\mu d\right)
=\frac{1}{3}\bar{q}\gamma^\mu v_\mu^{(s)}q + \bar{q}\gamma^\mu v_\mu q,
\end{equation}
from which we conclude
\begin{equation}
\label{extem}
v_\mu^{(s)}=-\frac{e}{2}{\cal A}_\mu,
\quad
v_\mu=-\frac{e}{2}\tau_3 {\cal A}_\mu,\quad
a_\mu=p=0,\quad
s=\hat{m} 1_{2\times 2}.
\end{equation}
   In this case the expressions for the divergence of the vector and
axial-vector currents, respectively, read
\begin{mathletters}
\begin{eqnarray}
\label{divvsc}
\partial_\mu V^\mu_i&=&-\epsilon_{3ij}e{\cal A}_\mu \bar{q}\gamma^\mu
\frac{\tau_j}{2}q=-\epsilon_{3ij}e{\cal A}_\mu V^\mu_j,\\
\label{divasc}
\partial_\mu A^\mu_i&=&-e {\cal A}_\mu \epsilon_{3ij} \bar{q}\gamma^\mu
\gamma_5 \frac{\tau_j}{2} q+\hat{m}i\bar{q}\gamma_5 \tau_i q
\nonumber\\
&=&-e {\cal A}_\mu \epsilon_{3ij} A^\mu_j+\hat{m}P_i,
\end{eqnarray}
\end{mathletters}
where we have introduced the isovector pseudoscalar density
\begin{equation}
\label{psd}
P_i=i\bar{q}\gamma_5 \tau_i q.
\end{equation}
   From Eq.\ (\ref{divasc}) we see that the axial-vector current is conserved,
if there is no external electromagnetic field {\em and}
if the quark mass vanishes (chiral limit).
  Strictly speaking, the right-hand side of Eq.\ (\ref{divasc}) should
also involve the anomaly term contributing to $\partial_\mu A^\mu_3$
\cite{Adler:1969gk,Adler:1969er,Bell:1969ts}.
   However, this term is of second order in the elementary charge and
thus not relevant for the below discussion of pion electroproduction which will
only be considered to first order in $e$.  
   We note the formal similarity of Eq.\ (\ref{divasc}) to the 
(pre-QCD) PCAC relation obtained by Adler through the inclusion of the 
electromagnetic interactions
with minimal electromagnetic coupling 
(see the appendix of Ref.\ \cite{Adler:1965}).\footnote{
   In Adler's version, the right-hand side of Eq.\ (\ref{divasc}) contains
a renormalized field operator creating and destroying pions.}
   Since in QCD the quarks are taken as truly elementary, their interaction
with an (external) electromagnetic field is of such a minimal type.

\subsection{Green functions}
   Before investigating the consequences of Eq.\ (\ref{divasc}) with respect
to the pion-electroproduction amplitude, we first have to define the 
nucleon matrix elements of the relevant quark bilinears and their
time-ordered products (Green functions).
   The first one is the nucleon matrix element of the axial-vector 
current,
\begin{equation}
\label{g1}
{\cal M}^\mu_{A, i}= \langle N(p_f) | A^\mu_i(0) | N(p_i) \rangle,
\end{equation}
   where the subscripts $A$ and $i$ refer to {\em axial-vector current} and
isospin component $i$, respectively.
   The matrix element of the time-ordered product of 
the electromagnetic current $J^\mu$ and the
isovector axial-vector current $A^\nu_i$ is defined as\footnote{Strictly
speaking one should work with the covariant time-ordered product ($T^\ast$)
which, typically, differs from the ordinary time-ordered product by
a non-covariant (seagull) term \cite{Jackiw:1972}.}
\begin{mathletters}
\label{g2}
\begin{eqnarray}
\label{g2a}
{\cal M}_{JA, i}^{\mu\nu}&=& 
\int d\,^4 x\, e^{-ik\cdot x} 
\langle N(p_f) | T\left[J^\mu(x) A^\nu_i(0)\right] | N(p_i) \rangle\\
\label{g2b}
&=&\int d\,^4 x\, e^{iq\cdot x}
\langle N(p_f) | T\left[J^\mu(0) A^\nu_i(x)\right] | N(p_i) \rangle.
\end{eqnarray}
\end{mathletters}
   Finally, the Green function involving the electromagnetic current 
$J^\mu$ and the isovector pseudoscalar density $P_i$ reads 
\begin{mathletters}
\label{g3}
\begin{eqnarray}
\label{g3a}
{\cal M}_{JP, i}^{\mu}
&=& \int d\,^4 x\, \,e^{-ik\cdot x} 
\langle N(p_f) | T\left[J^\mu(x) P_i(0) \right] | N(p_i) \rangle\\
\label{g3b}
&=& \int d\,^4 x\, \,e^{iq\cdot x} 
\langle N(p_f) | T\left[J^\mu(0) P_i(x) \right] | N(p_i) \rangle.
\end{eqnarray}
\end{mathletters}
   The Green functions of Eqs.\ (\ref{g1}) -- (\ref{g3}) are related
through the chiral Ward identity (for a pedagogical introduction to this 
topic see, e.g., Refs.\ \cite{Cheng:1984bj,Scherer:2002tk})\footnote{The 
standard derivation in terms of the ordinary time-ordered product leads 
to equal-time commutators of current densities and symmetry currents. 
   A naive application of the canonical commutation relations neglects 
the so-called Schwinger terms \cite{Schwinger:xd}.
   According to Feynman's conjecture these Schwinger terms cancel with
the above seagull terms.
   For a discussion of the validity of this hypothesis, the interested reader 
is referred to Ref.\ \cite{Jackiw:1972}.}
\begin{equation}
\label{gfrel}
q_\nu {\cal M}_{J A, i}^{\mu\nu}=
i\hat{m}{\cal M}_{J P, i}^\mu 
+\epsilon_{3ij}{\cal M}^\mu_{A,j}.
\end{equation}
   We will refer to Eq.\ (\ref{gfrel}) as the Adler-Gilman relation
(see Sec.\ II.B of Ref.\ \cite{Adler:1966} for their pre-QCD version).
   An alternative way of obtaining the analogue of 
Eq.\ (\ref{gfrel}) consists of evaluating the PCAC relation of Adler
\cite{Adler:1965},
\begin{equation}
(\partial_\mu \delta_{ij}+e {\cal A}_\mu \epsilon_{3ij})A^\mu_j=
m_\pi^2 F_\pi \Phi_i,
\end{equation}
between a final nucleon state and an initial state consisting of a
nucleon and a (virtual) photon (see, e.g., Refs.\
\cite{Vainshtein:ih,Dombey:1973ve,Scherer:1991cy,DeBaenst:1970hp,%
Ericson:1972,Naus:1990yj,Ohta:1992ui}).

\subsection{Interpolating field and pion electroproduction amplitude}

   The connection to the pion electroproduction amplitude is established
by noting that the pseudoscalar density serves as an interpolating pion 
field.
   The matrix element of the pseudoscalar quark density evaluated
between a single-pion state and the vacuum is defined in terms of
the coupling constant $G_\pi$ \cite{Gasser:1984yg},
\begin{equation}
\label{pipiv}
\langle 0|P_i(0)|\pi_j(q)\rangle = \delta_{ij}G_\pi.
\end{equation}
With the help of this constant we can define an interpolating pion
field\footnote{Every field $\Phi_i(x)$, which satisfies the relation
$\langle 0| \Phi_i(x)|\pi_j(q)\rangle = \delta_{ij} e^{-iq \cdot x}$,
can serve as an interpolating pion field.}
\begin{equation}
\label{pionfield}
  \Phi_i(x) = \frac{P_i(x)}{G_\pi}=\frac{P_i(x)}{2BF}=
  \frac{\hat{m} P_i(x)}{m_\pi^2F_\pi},
\end{equation}
where the second and third equality signs refer to the lowest-order
result of mesonic chiral perturbation theory 
\cite{Gasser:1984yg,Scherer:2002tk}.

   In the one-photon-exchange approximation, the invariant amplitude
for pion electroproduction\footnote{For a discussion of the relevant
kinematics and formalism see, e.g., Refs.\ 
\cite{Scherer:1991cy,Amaldi:1979vh}.}
 $\gamma^\ast(k)+N(p_i)\to \pi^i(q)+N(p_f)$
can be written as ${\cal M}_i=-ie \epsilon_\mu
{\cal M}^\mu_i$, where $\epsilon_\mu= e\bar{u}\gamma_\mu u/k^2$ is the
polarization vector of the virtual photon and ${\cal M}^\mu_i$
the transition-current matrix element:
\begin{equation}
\label{mmu}
{\cal M}^\mu_i=\langle N(p_f), \pi^i(q)|J^\mu(0)|N(p_i)\rangle.
\end{equation}
   Using the reduction formalism of Lehmann, Symanzik, and Zimmermann
(LSZ) \cite{Lehmann:1955rq} with the interpolating pion field of 
Eq.\ (\ref{pionfield}) in combination with the chiral Ward identity of 
Eq.\ (\ref{gfrel}), one obtains a relation for the transition-current 
matrix element in terms of QCD Green functions:
\begin{equation}
    \label{limgfrel}
{\cal M}^\mu_i = -i\frac{\hat{m}}{m_\pi^2F_\pi}\lim_{q^2\to m_\pi^2}
(q^2 - m_\pi^2) {\cal M}_{J P, i}^\mu
         = \frac{1}{m_\pi^2 F_\pi}\lim_{q^2\to m_\pi^2}(q^2 - m_\pi^2) 
(\epsilon_{3ij} {\cal M}^\mu_{A,j}-q_\nu {\cal M}^{\mu\nu}_{JA,i}).
\end{equation}
   This type of relation has been the starting point for studying the
consequences of the PCAC hypothesis on threshold photo- and electroproduction.
   Note that the QCD chiral Ward identity of Eq.\ (\ref{gfrel}) holds for any
value of $q^2$, i.e., there is no need to stay in the vicinity of the
squared pion mass, $q^2\approx m_\pi^2$.
   However, the connection to the physical pion-production
process requires $q^2=m_\pi^2$.

\section{The effective Lagrangian}
\label{section_effective_lagrangian}
   As already emphasized in the 1960's by Weinberg, phenomenological
Lagrangians provide a straightforward way of obtaining the results of current
algebra in the so-called phenomenological approximation, i.e., at tree level
\cite{Weinberg:1967fm,Weinberg:1968de,Gasiorowicz:1969kn,Dashen:1969ez}.
   Furthermore, modern techniques of effective field theory allow one
to also systematically calculate higher-order corrections to tree-level 
results in the framework of chiral perturbation theory 
\cite{Weinberg:1979kz,Gasser:1984yg,Gasser:1984gg,Gasser:1988rb,%
Weinberg:1991um,Jenkins:1990jv,Bernard:1992qa,Becher:1999he,Fuchs:2003qc}.

   In order to study the consequences of the chiral Ward identities for
threshold pion electroproduction, we will start from the most general effective
chiral Lagrangian up to and including ${\cal O}(p^3)$ in the baryonic sector.  
   However, for pedagogical purposes, we will make use of two (drastic)
simplifications. First, we will restrict ourselves to tree-level results only,
because they already reveal the main features regarding the connection 
between chiral Ward identities and pion production.
   The chiral corrections due to meson loops have been studied in detail
by Bernard {\em et al.} \cite{Bernard:1991rt,Bernard:1992ys,Bernard:1993rf}
and are beyond the scope of this work
(see Ref.\ \cite{Bernard:2001rs} for an overview).
   In particular, a discrepancy between the determination of the 
root-mean-square axial radius from anti-neutrino-proton scattering
and charged threshold pion-electroproduction, respectively, 
was explained in terms of such chiral pion-loop 
corrections \cite{Bernard:1992ys}.
   Second, even at the phenomenological level, we 
only consider a subset of terms which give rise to non-trivial contributions
to the Green functions.
   In the end, we will always comment on the full result at the one-loop
level.

\subsection{Mesonic Lagrangian}
   The most general lowest-order mesonic chiral Lagrangian can be 
written as \cite{Gasser:1984yg}
\begin{equation}
\label{l2}
{\cal L}_2 = \frac{F^2}{4} \mbox{Tr}\left[
D_\mu U (D^\mu U)^\dagger
+\chi U^\dagger + U\chi^\dagger
             \right],
\quad 
U(x)=\exp\left[i\frac{\vec{\tau}\cdot\vec{\pi}(x)}{F}\right],
\end{equation}
where the covariant derivative
\begin{equation}
\label{dmuu}
D_\mu U=\partial_\mu U-i r_\mu U+i U l_\mu
\end{equation}
involves the external fields of Eqs.\ (\ref{rtrans}) and (\ref{ltrans}),
and
\begin{equation}
\label{chi}
\chi=2 B(s+ip)
\end{equation}
contains the external scalar and pseudoscalar fields of Eqs.\ 
(\ref{spiptrans}) and (\ref{smiptrans}).
   Furthermore, $F$ is the pion-decay constant in the chiral limit,
$F_\pi = F\left[1 +{\cal O}(\hat{m})\right] = 92.4$ MeV, and
$B$ is related to the scalar quark condensate in the chiral limit,
$\langle 0 | \bar{u}u| 0\rangle_0 =
\langle 0 | \bar{d}d| 0\rangle_0 = -F^2B$.
   Inserting $s=\hat{m} 1_{2\times 2}$ in Eq.\ (\ref{l2}),
one finds, at lowest order in the momentum and quark-mass expansion, 
the relation $m_\pi^2=2B \hat{m}$.
   Combining Eqs.\ (\ref{localtrans}) with the transformation 
\begin{equation}
\label{utrans}
U\mapsto V_R U V_L^\dagger,
\end{equation}
the effective Lagrangian of Eq.\ (\ref{l2}) 
has the same {\em local} 
$\mbox{SU(2)}_L\times \mbox{SU(2)}_R\times \mbox{U}(1)_V$ symmetry
as the QCD Lagrangian of Eq.\ (\ref{lqcds}).

   According to the power counting scheme for processes involving
a single nucleon \cite{Weinberg:1979kz,Weinberg:1991um}, 
an ${\cal O}(p^3)$ calculation 
contains mesonic contributions of ${\cal O}(p^4)$.
   Thus, strictly speaking, we would also need the ${\cal O}(p^4)$ mesonic
Lagrangian of Gasser and Leutwyler \cite{Gasser:1984yg} for a full 
discussion. 
   However, we restrict ourselves to the ${\cal O}(p^2)$ contributions, 
because, in the above sense, they already illustrate
the relevant features regarding the chiral Ward identities.

\subsection{Pion-nucleon Lagrangian}
   The most general lowest-order relativistic pion-nucleon chiral Lagrangian 
reads \cite{Gasser:1988rb}
\begin{equation}
\label{lpin1}
{\cal L}^{(1)}_{\pi N}=\bar{\Psi}\left(i D\hspace{-.6em} / 
-m +
\frac{{\stackrel{\circ}{g}}_A}{2}
\gamma^\mu \gamma_5 u_\mu\right)\Psi,
\end{equation}
where
\begin{equation}
\label{psi}
\Psi=\left(\begin{array}{c} p\\ n\end{array}\right)
\end{equation}
is the nucleon isospin doublet.
   The covariant derivative is defined as 
\begin{equation}
\label{covderpsi}
D_\mu \Psi=\left(\partial_\mu +\Gamma_\mu -i v_\mu^{(s)}\right)\Psi,
\end{equation}
with
\begin{equation}
\label{Gammamu}
\Gamma_\mu= \frac{1}{2}\left[u^\dagger(\partial_\mu-i r_\mu)u
+u(\partial_\mu -i l_\mu)u^\dagger\right],
\end{equation}
and
\begin{equation}
\label{umu}
u_\mu=i\left[u^\dagger (\partial_\mu -i r_\mu)u
-u(\partial_\mu-i l_\mu)u^\dagger\right],
\end{equation}
where $u=\sqrt{U}$.
   At this order the effective Lagrangian contains two (new) low-energy 
constants, namely, the nucleon mass $m$
and the axial-vector coupling constant 
${\stackrel{\circ}{g}}_A$ in the chiral limit,
respectively.
   The Lagrangian of Eq.\ (\ref{lpin1}) is invariant
under {\em local} transformations, provided
\begin{equation}
\label{ntrans}
\Psi\mapsto \exp[-i\theta(x)](V_R U V_L^\dagger)^{-1/2} V_R \sqrt{U}\Psi,\quad
\quad U\mapsto V_R U V_L^\dagger.
\end{equation}
   We will neglect the next-to-leading-order pion-nucleon Lagrangian 
${\cal L}_{\pi N}^{(2)}$ \cite{Gasser:1988rb,Ecker:1996rk,Bernard:1997gq}, 
except for the terms giving rise to the isoscalar and isovector anomalous 
magnetic moments of the nucleon, respectively, 
\begin{equation}
\label{lpin2}
  {\cal L}_{\rm eff}^{(2)} = \frac{c_6}{2} 
          \bar{\Psi} \sigma^{\mu\nu} f^+_{\mu\nu}\Psi 
          + \frac{c_7}{2} \bar{\Psi} \sigma^{\mu\nu}
          \Psi (\partial_\mu v_\nu^{(s)} - \partial_\nu v_\mu^{(s)}),  
\end{equation}
where 
\begin{mathletters}
\label{flr}
\begin{eqnarray}
\label{fpm}
f^{\pm}_{\mu\nu}
&=& u f_{\mu\nu}^L u^\dagger \pm u^\dagger f_{\mu\nu}^R u , \\
\label{fr} 
f_{\mu\nu}^R&=&\partial_\mu r_\nu - \partial_\nu r_\mu- i[r_\mu, r_\nu], \\
\label{fl}
f_{\mu\nu}^L&=&\partial_\mu l_\nu - \partial_\nu l_\mu - i[l_\mu, l_\nu].
\end{eqnarray}
\end{mathletters}
   Inserting for the external fields the electromagnetic case of 
Eq.\ (\ref{extem}), the constants $c_6$ and $c_7$ are given in terms
of the isovector and isoscalar anomalous magnetic moments of the 
nucleon in the chiral limit, respectively, 
\begin{mathletters}
\label{kappa}
\begin{eqnarray}
\label{c6}
   c_6 &=& \frac{{\stackrel{\circ}{\kappa}}_v}{4 m},\\
\label{c7}
 \quad
   c_7 &=& \frac{{\stackrel{\circ}{\kappa}}_s}{2 m}.
\end{eqnarray}
\end{mathletters}
   To the order we are considering, all chiral limit values entering
Eq.\ (\ref{kappa}) can be replaced by their empirical values.
   The respective numerical values are 
$\kappa_s = \kappa_p + \kappa_n = -0.120$ and 
$\kappa_v =  \kappa_p - \kappa_n = 3.706$.
       
   In principle, the most general ${\cal L}^{(2)}_{\pi N}$ also
generates quark-mass corrections of ${\cal O}(\hat{m})$ to the nucleon mass 
and the axial-vector coupling constant, respectively. 
   The first one is not really important for our discussion, whereas the
second one will, in a similar fashion, also be provided by a 
${\cal O}(p^3)$ term, which for illustrative purposes we will keep.

   Finally, out of the $23$ terms of the third-order pion-nucleon Lagrangian
\cite{Ecker:1996rk}, re-written in relativistic notation, we only keep the 
following three terms,\footnote{In rewriting the heavy-baryon Lagrangian
of Ref.\ \cite{Ecker:1996rk} we made use of the replacement
$\bar{N}_v S_\mu N_v\mapsto \bar{\Psi}\gamma_\mu\gamma_5 \Psi/2$.
   We do not consider the $b_7$ and $b_8$ terms which generate a $q^2$ 
dependence of the electromagnetic form factors.
   For the discussion of the Adler-Gilman relation these terms do not
create any significant new features.}
\begin{equation}
   \label{lpin3}
   {\cal L}_{\rm eff}^{(3)} =\frac{1}{2(4\pi F)^2}
\bar{\Psi}\gamma^\mu\gamma_5
\left\{b_{17}u_\mu\mbox{Tr}(\chi_+)
+ib_{19}[D_\mu, \chi_{-}]
+b_{23}[D^\nu, f_{-\mu\nu}]\right\}\Psi
\end{equation}
with
\begin{equation}
\label{chipm}
\chi_{\pm} =  u^\dagger \chi u^\dagger \pm  u\chi^\dagger u.
\end{equation}
   The constants $b_{17}$, $b_{19}$, and $b_{23}$ will be discussed below.

   In order to summarize, in the following analysis we employ as the
effective Lagrangian
\begin{equation}
\label{leff}
{\cal L}_{\rm eff}={\cal L}_{2}+
{\cal L}^{(1)}_{\pi N}
+{\cal L}^{(2)}_{\rm eff}
+{\cal L}^{(3)}_{\rm eff},
\end{equation}
where the explicit expressions are given in Eqs.\ (\ref{l2}),
(\ref{lpin1}), (\ref{lpin2}), and (\ref{lpin3}).
   Within the tree-level approximation, it is consistent to
replace the nucleon mass in the chiral
limit, $m$, by the physical nucleon mass $m_N$.

\section{Axial-vector-current matrix elements}
\label{section_axial_vector-current_matrix_elements}

\subsection{First example}

   As the most simple application of the PCAC relation without an external
electromagnetic field, we first consider the matrix elements of the 
axial-vector current and the pseudoscalar density evaluated between a 
one-pion state and the vacuum.
   To that end, we insert the relevant external fields in the effective
Lagrangian and identify the vertices by applying the usual Feynman rules.
   This example serves as the most elementary illustration of how
chiral Ward identities are satisfied in our (simplified) approach  
and, more generally, in chiral perturbation theory.

\begin{figure}[ht]
   \begin{center}
   \mbox{\epsfig{file=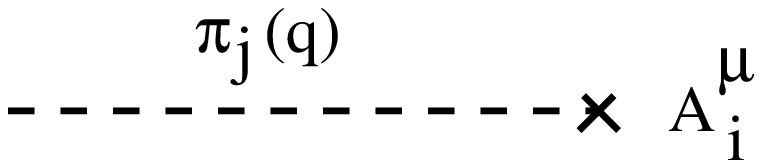,width=6.cm}}
   \end{center}
   \caption{\label{avcpion} Matrix element of the axial-vector current
   between a one-pion state and the vacuum.}
\end{figure}

\begin{figure}[ht]
   \begin{center}
   \mbox{\epsfig{file=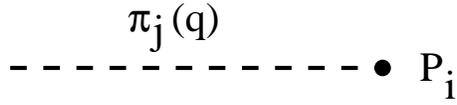,width=6.cm}}
   \end{center}
   \caption{\label{ppion} Matrix element of the pseudoscalar density
   between a one-pion state and the vacuum.}
\end{figure}

   Using Lorentz covariance and isospin symmetry, the matrix element of the 
axial-vector current can be parameterized as 
(see Fig.\ \ref{avcpion})
\begin{equation}
\label{axialcurrentpion}
\langle 0|A^\mu_i(x)|\pi_j(q)\rangle = i q^\mu F_\pi 
e^{-iq\cdot x}\delta_{ij}.
\end{equation}
   From the Lagrangian of Eq.\ (\ref{l2}) one obtains at ${\cal O}(p^2)$
\begin{equation}
\label{l2ext1}
{\cal L}_{\rm ext}
=i\frac{F^2}{2}\mbox{Tr}\left[
(\partial^\mu U U^\dagger -\partial^\mu U^\dagger U)
a_\mu\right]
=-F a_{\mu,i} \partial^\mu \pi_i+\cdots,
\end{equation}
which results in $F_\pi=F$ at this order.
   For the divergence of the axial vector current one then finds
\begin{eqnarray}
\langle 0|\partial_\mu A^\mu_i(x)|\pi_j(q)\rangle &=&
 i  q^\mu F_\pi \partial_\mu e^{-iq\cdot x}\delta_{ij}
=m^2_\pi F_\pi e^{-iq\cdot x}\delta_{ij}
= 2\hat{m} BF e^{-iq\cdot x}\delta_{ij},
\label{divaxialcurrentpion}
\end{eqnarray}
   where, in order to obtain the last equality sign, use has been made of the 
${\cal O}(p^2)$ predictions for $F_\pi$ and $m_\pi^2$, respectively.
   On the other hand, the matrix element of the pseudoscalar density
(see Fig.\ \ref{ppion}), Eq.\ (\ref{pipiv}), results from
\begin{equation}
\label{l2ext2}
{\cal L}_{\rm ext}
=i\frac{F^2B}{2}\mbox{Tr}(pU^\dagger-Up)=2BFp_i \pi_i+\cdots,
\end{equation}
   yielding $G_\pi = 2B F$. 
   Thus we have---at ${\cal O}(p^2)$---explicitly verified the relation
$F_\pi m_\pi^2=\hat{m} G_\pi$ implied by the PCAC relation. 
   The corresponding one-loop expressions can be found in 
Eqs.\ (12.4) and (12.6) of Ref.\ \cite{Gasser:1984yg}.
   
   Thus, within the framework of working in the phenomenological 
approximation of the effective Lagrangian of Eq.\ (\ref{leff}),
it is consistent to replace $F\to F_\pi$ and $2\hat{m} B\to m_\pi^2$.

\subsection{Nucleon matrix element of the pseudoscalar density}
\begin{figure}[ht]
   \begin{center}
   \mbox{\epsfig{file=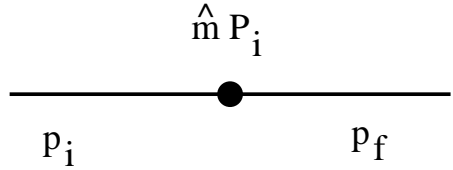, width=6.cm}}
   \end{center}
   \caption{\label{pnucleon} Matrix element of the pseudoscalar density
   between one-nucleon states.}
\end{figure}

\begin{figure}[ht]
   \begin{center}
   \mbox{\epsfig{file=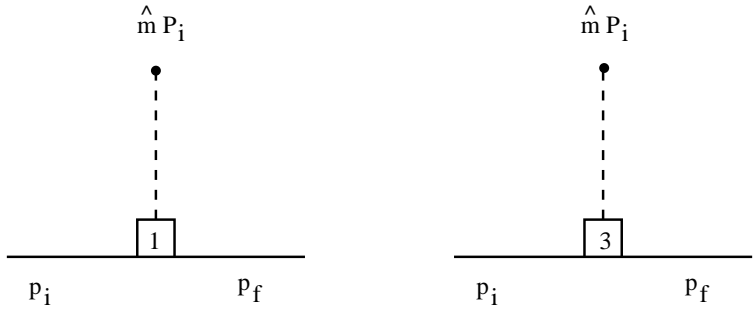,width=10.cm}}
   \end{center}
   \caption{\label{pnucleonfeynman} Feynman diagrams contributing to 
   the pseudoscalar density between one-nucleon states at ${\cal O}(p)$
   and ${\cal O}(p^3)$, respectively.}
\end{figure}

   The nucleon matrix element of the pseudoscalar density 
(see Fig.\ \ref{pnucleon})
can be parameterized as
\begin{equation}
\label{def_gt}
   \hat{m}\langle N(p_f)| P_i (0) | N(p_i) \rangle =
         \frac{m_\pi^2 F_\pi}{m_\pi^2 - t}\,
         G_{\pi N}(t)\,i\bar{u}(p_f) \gamma_5 \tau_i u(p_i),\quad
   t=(p_f-p_i)^2.
\end{equation}
   Since $\hat{m}P_i(x)/(m_\pi^2 F_\pi)$ serves as an interpolating pion field
[see Eq.\ (\ref{pionfield})], $G_{\pi N}$ 
is also referred to as the pion-nucleon
form factor (for this specific choice of interpolating field).
   In the framework of the effective chiral Lagrangian of Eq.\ (\ref{leff})
one obtains two contributions with vertices from ${\cal L}^{(1)}_{\pi N}$ 
and ${\cal L}^{(3)}_{\rm eff}$, respectively
(see Fig.\ \ref{pnucleonfeynman}),\footnote{At the order we are working
it is consistent to replace $F\to F_\pi$ and $m\to m_N$.}
\begin{equation}
\label{gt1}
G_{\pi N}(t)= \frac{m_N}{F_\pi}
\left({\stackrel{\circ}{g}}_A+\frac{b_{17}m^2_\pi}{4\pi^2 F_\pi^2}
-\frac{b_{19}}{8\pi^2 F_\pi^2 }t\right).
\end{equation}
   The pion-nucleon coupling constant is defined at $t=m_\pi^2$:
\begin{equation}
\label{gpinn}
g_{\pi N}=G_{\pi N}(m_\pi^2)=
\frac{m_N}{F_\pi}\left({\stackrel{\circ}{g}}_A
+\frac{(2 b_{17}-b_{19}) m_\pi^2}{8\pi^2 F_\pi^2}
\right).
\end{equation}
   As will be seen below, 
$g_A={\stackrel{\circ}{g}}_A+b_{17}m^2_\pi/(4\pi^2 F_\pi^2)$,
such that the parameter $b_{19}$ reflects the so-called 
Goldberger-Treiman discrepancy, i.e., the numerical violation of the 
Goldberger-Treiman relation,\footnote{
Using $m_N=938.3$ MeV, $g_A=1.267$, $F_\pi=92.4$ MeV, 
and $g_{\pi N}=13.21$  \cite{Schroder:rc},
one obtains $\Delta_{\pi N}=2.6$ \%.}
\begin{equation}
\label{gtd}
\Delta_{\pi N}\equiv 
1-\frac{g_A m_N}{g_{\pi N} F_\pi}=-b_{19}\frac{m_N m_\pi^2}{
8\pi^2 g_{\pi N} F_\pi^3}.
\end{equation}
    The effective Lagrangian of Eq.\ (\ref{leff}) reproduces exactly
the same result for $\Delta_{\pi N}$ as the full one-loop calculation
[see Eq.\ (69) of Ref.\ \cite{Fearing:1997dp}].

   In terms of $\Delta_{\pi N}$ and $g_{\pi N}$, the pion-nucleon form
factor---at this order---can be rewritten as
\begin{equation}
\label{gt2}
G_{\pi N}(t)= g_{\pi N}\left(1+\Delta_{\pi N} \frac{t-m^2_\pi}{m_\pi^2}\right).
\end{equation}
   The full ${\cal O}(p^3)$ calculation including pion loop corrections
generates exactly the same functional form with all quantities
replaced by their ${\cal O}(p^3)$ expressions (see Sec.\ III.7
of Ref.\ \cite{Bernard:1995dp}). 

\subsection{The pion-nucleon vertex}
   The pion-nucleon vertex resulting from Eq.\ (\ref{leff}) reads
\begin{equation}
\label{pnvertex}
\left({\stackrel{\circ}{g}}_A+\frac{b_{17}m_\pi^2}{4\pi^2 F_\pi^2}
-\frac{b_{19}m_\pi^2}{8\pi^2 F_\pi^2 }\right)\frac{1}{F_\pi}
q\hspace{-.5em}/\hspace{.3em} \gamma_5\frac{\tau_i}{2}
=\frac{g_{\pi N}}{2 m_N} q\hspace{-.5em}/\hspace{.3em} \gamma_5 \tau_i,\quad
q=p_i-p_f,
\end{equation}
where we made use of Eq.\ (\ref{gpinn}).
   In particular, for $q^2\neq m_\pi^2$ the pion-nucleon vertex does
{\em not} contain the form factor $G_{\pi N}(q^2)$ of Eq.\ (\ref{def_gt}).   
   In general, the pion-nucleon vertex depends on the choice of the field
variables in the (effective) Lagrangian.  
   In the present case, the pion-nucleon vertex is only an auxiliary
quantity, whereas the ``fundamental'' quantity (entering chiral Ward
identities) is the QCD Green function of Eq.\ (\ref{def_gt}).
      Only at $q^2=m_\pi^2$, we expect the same coupling strength, since
both $\Phi_i$ of Eq.\ (\ref{pionfield}) and $\pi_i$ of (\ref{l2}) 
serve as interpolating pion fields.

\subsection{Nucleon axial-vector-current matrix element}
\begin{figure}[ht]
   \begin{center}
   \mbox{\epsfig{file=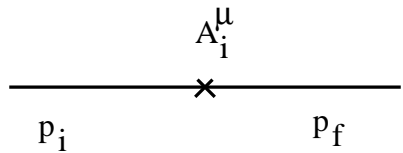, height=2.cm}}
   \end{center}
   \caption{\label{avcnucleon} Matrix element of the axial-vector current
   between one-nucleon states.}
\end{figure}

\begin{figure}[ht]
   \begin{center}
   \mbox{\epsfig{file=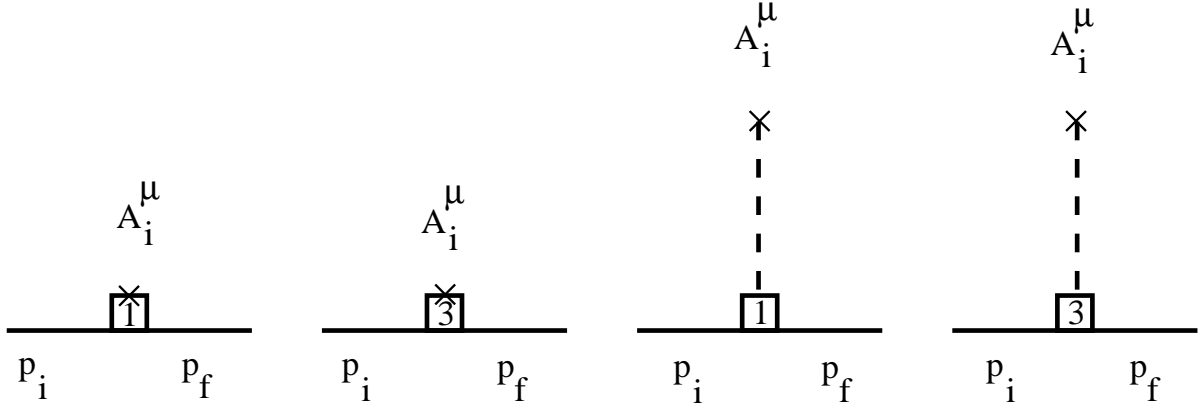,width=16.cm}}
   \end{center}
   \caption{\label{avcnucleonfeynman} Feynman diagrams contributing to 
   the axial-vector current matrix element between one-nucleon states at 
   ${\cal O}(p)$ and ${\cal O}(p^3)$.}
\end{figure}

   We now turn to the results for the axial form 
factors of the nucleon.\footnote{For simplicity, we often refer to {\em both}
form factors parameterizing the axial-vector-current matrix element 
as axial form factors.}
   The matrix element of the axial-vector current
evaluated between initial and final nucleon states---excluding second-class 
currents \cite{Weinberg:1958ut}---can be written as 
(see Fig.\ \ref{avcnucleon})
\begin{equation}
\label{axial_current}
   \left<N(p_f)| A_i^\mu (0) | N(p_i) \right> =
          \bar{u}(p_f) \left[ \gamma^\mu G_A(t) + \frac{(p_f-p_i)^\mu}{2m_N}
          G_P(t) \right] \gamma_5 \frac{\tau_i}{2} u(p_i),
\end{equation}
where $t = (p_f-p_i)^2$, and $G_A(t)$ and $G_P(t)$ are the 
axial and induced pseudoscalar form factors, respectively.
   Within the framework of Eq.\ (\ref{leff}) we obtain to this order (see
Fig.\ \ref{avcnucleonfeynman})
\begin{mathletters}
\label{gagp}
\begin{eqnarray}
\label{ga}
   G_A(t) &=& g_A \left( 1 + \frac{1}{6} \ra\, t \right),\\
\label{gp}
   G_P(t) &=& 4m_N^2\left( \frac{F_\pi g_{\pi N}}{m_N} \frac{1}{m_\pi^2 -t}
              - \frac{1}{6} g_A \ra \right),
\end{eqnarray}
\end{mathletters}
   where 
\begin{mathletters}
\label{axpar}
\begin{eqnarray}
\label{axcoup}
g_A=G_A(0)&=&{\stackrel{\circ}{g}}_A+\frac{b_{17}m^2_\pi}{4\pi^2 F_\pi^2},\\
\label{axrad}
-\frac{1}{6}g_A\langle r^2\rangle_A&=&\frac{b_{23}}{(4\pi F_\pi)^2}.
\end{eqnarray}
\end{mathletters}
   In the present framework, the parameter $b_{17}$ signifies a deviation
of the axial-vector coupling constant $g_A$ from its value
${\stackrel{\circ}{g}}_A$  in the chiral limit.
   The parameter $b_{23}$ is related to the axial radius.   
   The full one-loop calculation at ${\cal O}(p^3)$ 
\cite{Fearing:1997dp} has the same functional form as Eqs.\ (\ref{gagp}).
   In addition to the $b_{17}$ term, the pion loops generate a further 
contribution to $g_A=G_A(0)$ of order $\hat{m}$ and $\hat{m}\ln(\hat{m})$.
   Its infinite piece is compensated by an infinity in the {\em bare} parameter
$b_{17}^0$.

   It is now straightforward to verify that the form factors of Eqs.\
(\ref{gt2}) and (\ref{gagp}) satisfy the relation 
\begin{equation}
\label{ff_relation}
   2m_N G_A(t) + \frac{t}{2m_N} G_P(t) =
       2\frac{m_\pi^2 F_\pi}{m_\pi^2 - t} G_{\pi N}(t),
\end{equation}
as implied by the PCAC relation, Eq.\ (\ref{divasc}), in the absence
of an electromagnetic field ${\cal A}_\mu$.
   Evaluating Eq.\ (\ref{ff_relation}) at $t=0$, one obtains
\begin{equation}
\label{ff_rel_t0}
2 m_N G_A(0) = 2F_\pi G_{\pi N}(0),
\end{equation}
   where we made use of the fact that $G_P(0)$ is finite for nonvanishing
$m_\pi^2$.
   By use of $G_{\pi N}(0)=g_{\pi N}(1-\Delta_{\pi N}) \approx g_{\pi N}
=G_{\pi N}(m_\pi^2)$, we see that the Goldberger-Treiman relation is only
approximately satisfied.
  Of course, in the chiral limit we recover 
\begin{equation}
2 m \stackrel{\circ}{G}_A(t)+\frac{t}{2 m}
\stackrel{\circ}{G}_P(t)=0,
\end{equation}
   which also implies that the Goldberger-Treiman relation is exactly 
satisfied in this case.

\section{Pion electroproduction}
\label{section_pion_electroproduction}
   We will now address with the help of ${\cal L}_{\rm eff}$ of Eq.\
(\ref{leff}) how the PCAC relation enters the pion electroproduction 
amplitude.
   Neglecting isospin-symmetry-breaking effects due to different $u$- and
$d$-quark masses as well as higher-order electromagnetic corrections, the
amplitude for producing a pion with Cartesian isospin index $i$ can
be decomposed as \cite{Chew:1957tf}
\begin{equation}
\label{isospindecomposition}
{\cal M}(\pi_i)=\chi_f^\dagger\left(-i\epsilon_{3ij}\tau_j {\cal M}^{(-)}
+\tau_i {\cal M}^{(0)}+\delta_{i3} {\cal M}^{(+)}\right)\chi_i,
\end{equation}
where $\chi_i$ and $\chi_f$ denote the isospinors of the initial and
final nucleons, respectively, and $\tau_i$ are the ordinary Pauli
matrices.
   With this decomposition the amplitudes for the physical processes read
\begin{mathletters}
\label{physchan}
\begin{eqnarray}
\label{mpiplus}
{\cal M}(\gamma^\ast p\to n\pi^+)&=&\sqrt{2}({\cal M}^{(0)}+{\cal M}^{(-)}),\\
\label{mpiminus}
{\cal M}(\gamma^\ast n\to p\pi^-)&=&\sqrt{2}({\cal M}^{(0)}-{\cal M}^{(-)}),\\
\label{mppi0}
{\cal M}(\gamma^\ast p\to p\pi^0)&=&{\cal M}^{(+)}+{\cal M}^{(0)},\\
\label{mnpi0}
{\cal M}(\gamma^\ast n\to n\pi^0)&=&{\cal M}^{(+)}-{\cal M}^{(0)}.
\end{eqnarray}
\end{mathletters}

\subsection{Direct calculation}
\begin{figure}[ht]
   \begin{center}
   \mbox{\epsfig{file=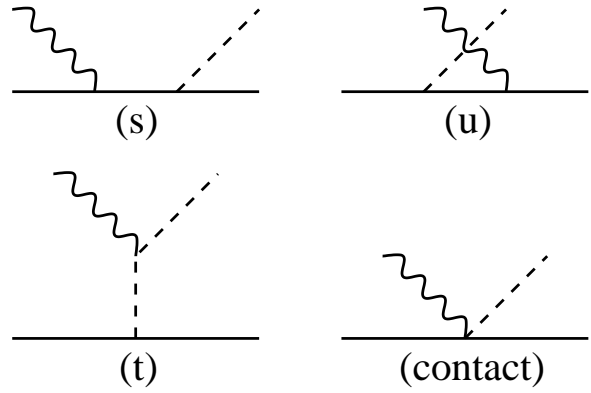, height=5cm}}
   \end{center}
   \caption{\label{pionelectroproductiondiagrams} 
Feynman diagrams contributing to
pion electroproduction in the framework of ${\cal L}_{\rm eff}$ of
Eq.\ (\ref{leff}).}
\end{figure}

   The natural way to find the pion electroproduction amplitude
associated with the effective Lagrangian of Eq.\ (\ref{leff}) is to
determine the relevant vertices involving pions, nucleons, and the
electromagnetic field and to calculate the corresponding Feynman diagrams.
   The calculation is straightforward and involves the diagrams shown in 
Fig.\ \ref{pionelectroproductiondiagrams}:
\begin{mathletters}
\label{mpepd}
\begin{eqnarray}
\label{ms}
{\cal M}_s&=&-e\frac{g_{\pi N}}{m_N}\bar{u}(p_f)
\left(1-\frac{2 m_N q\hspace{-.5em}/}{s-m_N^2}\right)\gamma_5
\frac{\tau_i}{2}\epsilon\cdot\Gamma(k)u(p_i),\\
\label{mu}
{\cal M}_u&=&-e\frac{g_{\pi N}}{m_N}\bar{u}(p_f)
\epsilon\cdot\Gamma(k)
\left(1-\frac{2 m_N q\hspace{-.5em}/}{u-m_N^2}\right)
\gamma_5 \frac{\tau_i}{2}u(p_i),\\
\label{mt}
{\cal M}_t&=&ie g_{\pi N}\epsilon_{3ij}\tau_j\bar{u}(p_f)\gamma_5
u(p_i)\frac{1}{t-m_\pi^2} \epsilon\cdot(2q-k),\\
\label{mcontact}
{\cal M}_{\rm contact}&=&
ie\epsilon_{3ij}\tau_j \frac{g_{\pi N}}{2m_N}\bar{u}(p_f)
\epsilon\hspace{-.45em}/\hspace{.3em}\gamma_5 u(p_i)\nonumber\\
&&+ie\epsilon_{3ij}\tau_j \frac{g_A}{2 F_\pi}
\frac{1}{6}\langle r^2\rangle_A
\bar{u}(p_f)[(k-q)\cdot k \epsilon\hspace{-.45em}/\hspace{.3em}
-(k-q)\cdot\epsilon k\hspace{-.5em}/]\gamma_5 u(p_i),
\end{eqnarray}
\end{mathletters}     
   where $s=(p_i+k)^2$, $t=(p_i-p_f)^2$, and $u=(p_i-q)^2$ 
are the usual Mandelstam variables,
satisfying $s+t+u=2m_N^2+k^2+m_\pi^2$.\footnote{In the case of an
off-shell pion one has to replace $m_\pi^2$ by $q^2$.} 
  The expressions for $g_{\pi N}$, $g_A$, and $\langle r^2\rangle_A$ are 
given in Eqs.\ (\ref{gpinn}), (\ref{axcoup}), and (\ref{axrad}), respectively. 
   Furthermore, we introduced the abbreviation 
\begin{equation}
\label{Gamma}
\Gamma^\mu(k)=\gamma^\mu \frac{1+\tau_3}{2}
+i\frac{\sigma^{\mu\nu}k_\nu}{2m_N}
\left(\frac{\kappa_s}{2}+\frac{\kappa_v}{2}\tau_3\right),
\quad k=p_f-p_i,
\end{equation}
for the electromagnetic vertex of the nucleon (as obtained in the framework
of the effective Lagrangian).
   First of all, it is straightforward to check the constraints of 
gauge invariance for Eqs.\ (\ref{mpepd}) \cite{Kazes_1959}:
\begin{equation}
\label{sgipep}
k_\mu {\cal M}^\mu_i=-g_{\pi N} \epsilon_{3ij}\tau_j
\bar{u}(p_f)\gamma_5 u(p_i) \Delta^{-1}(q) \Delta (q-k).
\end{equation}
   Equation (\ref{sgipep}) is the electromagnetic Ward-Takahashi identity for 
the production of an off-shell pion consistent with the vertices and 
propagators obtained from ${\cal L}_{\rm eff}$ 
[see, e.g., Eq.\ (4) of Ref.\ \cite{Bos:1992qs}],
where the external nucleon lines are on shell.
   This result is not surprising, because the transformation law
of Eq.\ (\ref{utrans}) for the chiral matrix $U$ implies
for the pion field 
$\pi_i\mapsto \pi_i-\theta(x)\epsilon_{3ij}\pi_j$  
under electromagnetic U(1) transformations.
   In particular, if the pion is on its mass shell, $q^2=m_\pi^2$,
one obtains the usual current conservation condition,
$k_\mu {\cal M}^\mu_i=0$, because $\Delta^{-1}(q)=0$ in this case.

   At this point we can now point out the distinction between chiral Ward 
identities relating QCD Green functions and (electromagnetic) 
Ward-Takahashi identities relating Green functions
of the effective theory containing off-shell legs of the effective
degrees of freedom, here pions and nucleons.  
   The chiral Ward identities originate in the chiral symmetry of the
underlying QCD Lagrangian.
   By considering the (most general) chiral effective Lagrangian 
exhibiting the same invariances as the QCD Lagrangian coupled to external
fields, the constraints of the chiral Ward identities are automatically
transported to the effective-Lagrangian level.
   On the other hand, the effective degrees of freedom are carriers of,
e.g., U(1) representations resulting, in addition, in conventional 
Ward-Takahashi identities involving off-shell pion and nucleon 
vertices.\footnote{Of course, 
also other groups which are linearly
realized on the pion and nucleon degrees of freedom, such as SU(2)$_V$,
may be used to obtain consistency checks between the building blocks of 
the effective theory.}
   An example of such a Ward-Takahashi identity is given by 
Eq.\ (\ref{sgipep}).
   Note, that neither the left-hand nor the right-hand side constitute
QCD Green functions. 

   Finally, Eq.\ (\ref{mcontact}) which, according to Eqs.\ (\ref{physchan}),
only enters charged pion production, involves the axial radius.
   In fact, this is not a coincidence, but will be shown to also follow
from the (more complicated) application of the Adler-Gilman relation.

\subsection{Pion electroproduction and the electromagnetic-current 
pseudoscalar-density Green function}

\begin{figure}[ht]
   \begin{center}
   \mbox{\epsfig{file=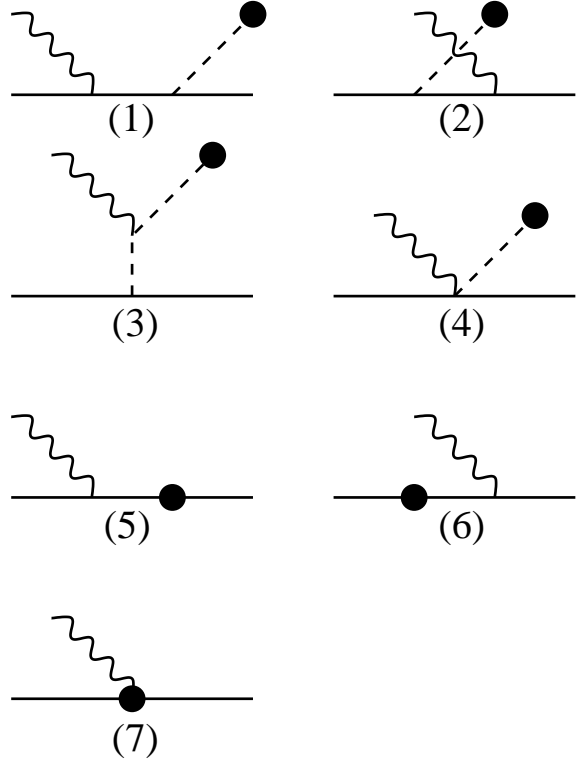, height=10cm}}
   \end{center}
   \caption{\label{mmujpifig} Feynman diagrams contributing to
the Green function involving the electromagnetic current and
the pseudoscalar density.  
The wavy line denotes a (virtual) photon coupling to the electromagnetic
current. 
   The full circle corresponds to the pseudoscalar density.}
\end{figure}

   According to Eq.\ (\ref{limgfrel}), the pion electroproduction transition 
current matrix element is related to the QCD Green function involving
the electromagnetic current and the pseudoscalar density. 
   Here, we calculate ${\cal M}^\mu_{JP,i}$ of Eq.\ (\ref{g3}) in
the framework of ${\cal L}_{\rm eff}$ and explicitly verify the result for 
the pion electroproduction amplitude of Eqs.\ (\ref{mpepd}). 
   The result is obtained from the seven diagrams shown in 
Fig.\ \ref{mmujpifig}:
\begin{mathletters}
\label{mmujpi}
\begin{eqnarray}
\label{mmujpi1}
{\cal M}^\mu_{JP,i,1}&=&\frac{2B F_\pi}{q^2-m_\pi^2}\frac{g_{\pi N}}{m_N}
\bar{u}(p_f)\left(1-\frac{2 m_N q\hspace{-.5em}/}{s-m_N^2}\right)\gamma_5
\frac{\tau_i}{2}\Gamma^\mu(k)u(p_i),\\
\label{mmujpi2}
{\cal M}^\mu_{JP,i,2}&=&\frac{2B F_\pi}{q^2-m_\pi^2}\frac{g_{\pi N}}{m_N}
\bar{u}(p_f)\Gamma^\mu(k)
\left(1-\frac{2 m_N q\hspace{-.5em}/}{u-m_N^2}\right)\gamma_5\frac{\tau_i}{2} 
 u(p_i),\\
\label{mmujpi3}
{\cal M}^\mu_{JP,i,3}&=&-i\epsilon_{3ij}\tau_j
\frac{2B F_\pi}{q^2-m_\pi^2}
\frac{1}{t-m_\pi^2}(2q-k)^\mu g_{\pi N}\bar{u}(p_f)\gamma_5  u(p_i),\\
\label{mmujpi4}
{\cal M}^\mu_{JP,i,4}&=&-i\epsilon_{3ij}\tau_j
\frac{2B F_\pi}{q^2-m_\pi^2}
\bar{u}(p_f)\left\{\frac{g_{\pi N}}{2m_N}\gamma^\mu\right.\nonumber\\
&&\left.
+\frac{1}{6}\langle r^2\rangle_A \frac{g_A}{2F_\pi}\left[
(k-q)\cdot k \gamma^\mu- (k-q)^\mu k\hspace{-.5em}/\right]\right\}\gamma_5
u(p_i),\\
\label{mmujpi5}
{\cal M}^\mu_{JP,i,5}&=&\frac{2B F_\pi}{m^2_\pi}\Delta_{\pi N}
\frac{g_{\pi N}}{m_N}
\bar{u}(p_f)\left(1-\frac{2 m_N q\hspace{-.5em}/}{s-m_N^2}\right)\gamma_5
\frac{\tau_i}{2}\Gamma^\mu(k)u(p_i),\\
\label{mmujpi6}
{\cal M}^\mu_{JP,i,6}&=&\frac{2B F_\pi}{m^2_\pi}\Delta_{\pi N} 
\frac{g_{\pi N}}{m_N}
\bar{u}(p_f)\Gamma^\mu(k)
\left(1-\frac{2 m_N q\hspace{-.5em}/}{u-m_N^2}\right)\gamma_5
\frac{\tau_i}{2}  u(p_i),\\
\label{mmujpi7}
{\cal M}^\mu_{JP,i,7}&=&-i\epsilon_{3ij}\tau_j 
\frac{2B F_\pi}{m^2_\pi}\Delta_{\pi N} \frac{g_{\pi N}}{2m_N}
\bar{u}(p_f)\gamma^\mu\gamma_5 u(p_i).
\end{eqnarray}
\end{mathletters}
   The expression of the Goldberger-Treiman discrepancy is given in 
Eq.\ (\ref{gtd}).
   When multiplying Eqs.\ (\ref{mmujpi}) by $-\hat{m} i$ we make use of
$2\hat{m}B F_\pi=m_\pi^2 F_\pi$. 
   Secondly, after multiplying by $q^2-m_\pi^2$ and taking the limit
$q^2\to m_\pi^2$, only those terms of Eqs.\ (\ref{mmujpi})
which have a $1/(q^2-m_\pi^2)$ singularity
survive. Finally, in order to obtain the invariant amplitude we
have to contract the result with $-ie \epsilon_\mu$. 
   With these replacements one easily sees the one-to-one correspondence
between Eqs.\ (\ref{ms}) - (\ref{mcontact}) and Eqs.\ (\ref{mmujpi1})
- (\ref{mmujpi4}).  
   On the other hand Eqs.\ (\ref{mmujpi5}) - (\ref{mmujpi7}) do not
contribute to pion electroproduction due to the absence of the
$1/(q^2-m_\pi^2)$ pole.
   Thus, we have a first check of the consistency of our procedure.

   As a final check of the results of Eqs.\ (\ref{mmujpi}) we investigate
the chiral Ward identity
\begin{equation}
\label{cwikmummujpi}
k_\mu {\cal M}^\mu_{JP,i}=\epsilon_{3ij}\langle N(p_f)|P_j(0)|N(p_i)\rangle.
\end{equation}
   Contracting the first four and the final three expressions of 
Eqs.\ (\ref{mmujpi}) with $k_\mu$, respectively, we obtain
\begin{mathletters}
\label{contract}
\begin{eqnarray*}
k_\mu \sum_{k=1}^4 {\cal M}^\mu_{JP,i,k}&=&-i\epsilon_{3ij}\tau_j
\frac{2B F_\pi}{t-m_\pi^2} g_{\pi N} \bar{u}(p_f)\gamma_5 u(p_i),\\
k_\mu \sum_{k=5}^7 {\cal M}^\mu_{JP,i,k}&=&-i\epsilon_{3ij}\tau_j
\frac{2B F_\pi}{m_\pi^2}\Delta_{\pi N} g_{\pi N} \bar{u}(p_f)\gamma_5 u(p_i).
\end{eqnarray*}
\end{mathletters}
   Combining the two results we find
\begin{equation}
\label{cwikmummujpiexpl}
k_\mu {\cal M}^\mu_{JP,i}=-i\epsilon_{3ij}\tau_j \frac{2B F_\pi}{t-m_\pi^2}
G_{\pi N}(t) \bar{u}(p_f)\gamma_5 u(p_i),
\end{equation}
where $G_{\pi N}(t)$ is given in Eq.\ (\ref{gt2}).
   Here, we made use of the definition of Eq.\ (\ref{def_gt}) and 
$2\hat{m} B =m_\pi^2$.
   Thus, the result for the Green function ${\cal M}^\mu_{JP,i}$
is consistent with the chiral Ward identity of Eq.\ (\ref{cwikmummujpi}).

\subsection{Adler-Gilman relation}

   We now turn to the explicit test of the
Adler-Gilman relation, Eq.\ (\ref{gfrel}), in the
framework of ${\cal L}_{\rm eff}$. 
   In traditional current-algebra or soft-pion approaches, it is the 
right-hand-side of Eq.\ (\ref{gfrel}) which serves as the starting point 
for the prediction of threshold pion production.

\subsubsection{Electromagnetic-current axial-vector current Green function}  
   We first need to calculate the Green function 
${\cal M}^{\mu\nu}_{JA,i}$ of Eq.\ (\ref{g2}) involving
the electromagnetic current and the axial-vector current 
(see Fig.\ \ref{mmunujaifig}):
\begin{mathletters}
\label{mmunujai}
\begin{eqnarray}
\label{mmunujai1}
{\cal M}^{\mu\nu}_{JA,i,1}
&=&i\frac{F_\pi q^\nu}{q^2-m_\pi^2}\frac{g_{\pi N}}{m_N}
\bar{u}(p_f)\left(1-\frac{2 m_N q\hspace{-.5em}/}{s-m_N^2}\right)\gamma_5
\frac{\tau_i}{2}\Gamma^\mu(k)u(p_i),\\
\label{mmunujai2}
{\cal M}^{\mu\nu}_{JA,i,2}
&=&
i\frac{F_\pi q^\nu}{q^2-m_\pi^2}\frac{g_{\pi N}}{m_N}
\bar{u}(p_f)\Gamma^\mu(k)
\left(1-\frac{2 m_N q\hspace{-.5em}/}{u-m_N^2}\right)
\gamma_5\frac{\tau_i}{2} 
 u(p_i),\\
\label{mmunujai3}
{\cal M}^{\mu\nu}_{JA,i,3}
&=&\epsilon_{3ij}\tau_j \frac{F_\pi q^\nu}{q^2-m_\pi^2}
\frac{(2q-k)^\mu}{t-m_\pi^2} g_{\pi N}\bar{u}(p_f)\gamma_5  u(p_i),\\
\label{mmunujai4}
{\cal M}^{\mu\nu}_{JA,i,4}
&=&
\epsilon_{3ij}\tau_j
\frac{F_\pi q^\nu}{q^2-m_\pi^2}
\bar{u}(p_f)\left\{\frac{g_{\pi N}}{2m_N}\gamma^\mu\right.\nonumber\\
&&\left.
+\frac{1}{6}\langle r^2\rangle_A \frac{g_A}{2F_\pi}\left[
(k-q)\cdot k \gamma^\mu- (k-q)^\mu k\hspace{-.5em}/\hspace{.5em}
\right]\right\}\gamma_5
u(p_i),\\
\label{mmunujai5}
{\cal M}^{\mu\nu}_{JA,i,5}
&=&i g_A \frac{1}{s-m_N^2}
\bar{u}(p_f)\left[\gamma^\nu\left(1+\frac{1}{6}\langle r^2\rangle_A\, 
q^2\right)
-q^\nu q\hspace{-.5em}/\hspace{.5em}
\frac{1}{6}\langle r^2\rangle_A\right]
\gamma_5 \frac{\tau_i}{2}\nonumber\\
&&\times (p\hspace{-.5em}/\hspace{.1em}_i
+k\hspace{-.5em}/ +m_N) \Gamma^\mu(k)u(p_i),\\
\label{mmunujai6}
{\cal M}^{\mu\nu}_{JA,i,6}
&=&
i g_A \frac{1}{u-m_N^2}
\bar{u}(p_f)\Gamma^\mu(k)(p\hspace{-.5em}/\hspace{.1em}_i
-q\hspace{-.5em}/ +m_N)\nonumber\\
&&\times
\left[\gamma^\nu\left(1+\frac{1}{6}\langle r^2\rangle_A \,q^2\right)
-q^\nu q\hspace{-.5em}/\hspace{.5em}
\frac{1}{6}\langle r^2\rangle_A\right]
\gamma_5 \frac{\tau_i}{2}u(p_i),\\
\label{mmunujai7}
{\cal M}^{\mu\nu}_{JA,i,7}
&=&
-\epsilon_{3ij}\tau_j g_A \frac{1}{6}\langle r^2\rangle_A
\frac{1}{2}
\bar{u}(p_f)
\left[\gamma^\nu(2q-k)^\mu-\gamma^\mu(q-k)^\nu-q\hspace{-.5em}/\hspace{.5em}
g^{\mu\nu}\right]\gamma_5 u(p_i),\\
\label{mmunujai8}
{\cal M}^{\mu\nu}_{JA,i,8}
&=&-\epsilon_{3ij}\tau_j g_{\pi N} F_\pi \frac{g^{\mu\nu}}{t-m_\pi^2}
\bar{u}(p_f)\gamma_5 u(p_i).
\end{eqnarray}
\end{mathletters}
   Note that Eqs.\ (\ref{mmunujai1}) - (\ref{mmunujai4}) are obtained
from Eqs.\ (\ref{mmujpi1}) - (\ref{mmujpi4}) by the replacement
$2 B\to iq^\nu$ which simply reflects the respective coupling of the 
external pseudoscalar and the axial-vector fields to a single pion 
resulting from Eq.\ (\ref{l2}).
   Moreover, the coupling to the axial-vector current provides the additional
term of Eq.\ (\ref{mmunujai8}) in comparison with ${\cal M}^\mu_{JP,i}$
of Eq.\ (\ref{mmujpi}).

\subsubsection{Gauge invariance}
   As a first test of the results of Eqs.\ (\ref{mmunujai}) we will 
investigate electromagnetic gauge invariance by contracting  
${\cal M}^{\mu\nu}_{JA,i}$ with the four-momentum $k_\mu$.
   The corresponding chiral Ward identity reads
\begin{equation}
\label{cwikmummunujai}
k_\mu {\cal M}^{\mu\nu}_{JA,i}
=\epsilon_{3ij}\langle N(p_f)|A^\nu_j(0)|N(p_i)\rangle.
\end{equation}
   Contracting the sum of the first four and the final expressions,
and the sum of the remaining three, respectively, with $k_\mu$ we obtain
\begin{mathletters}
\label{contract2}
\begin{eqnarray}
k_\mu \sum_{k=1}^4 {\cal M}^{\mu\nu}_{JA,i,k}
+k_\mu {\cal M}^{\mu\nu}_{JA,i,8}
&=&-\epsilon_{3ij}\tau_j
\frac{p_f^\nu-p_i^\nu}{t-m_\pi^2} g_{\pi N} F_\pi 
\bar{u}(p_f)\gamma_5 u(p_i),\\
k_\mu \sum_{k=5}^7 {\cal M}^{\mu\nu}_{JA,i,k}&=&
\epsilon_{3ij} g_A \bar{u}(p_f)
\left[\gamma^\nu\left(1+\frac{1}{6}\langle r^2\rangle_A\, t\right)\right.
\nonumber\\
&&\left.
-2m_N (p_f-p_i)^\nu\frac{1}{6}
\langle r^2 \rangle_A \right]\gamma_5 \frac{\tau_j}{2}u(p_i).
\end{eqnarray}
\end{mathletters}
   Adding the two terms and comparing with the result for the 
axial-vector current matrix element of Eqs.\ (\ref{axial_current})
and (\ref{gagp}), we see that the chiral Ward identity
of Eq.\ (\ref{cwikmummunujai}) is indeed satisfied.

\subsubsection{Test of the Adler-Gilman relation}
   We are finally in the position to explicitly test the 
Adler-Gilman relation. 
Since this has been the key ingredient in many investigations of the
connection between the PCAC hypothesis and threshold pion production,
we will discuss the individual terms in detail.

\begin{figure}[ht]
   \begin{center}
   \mbox{\epsfig{file=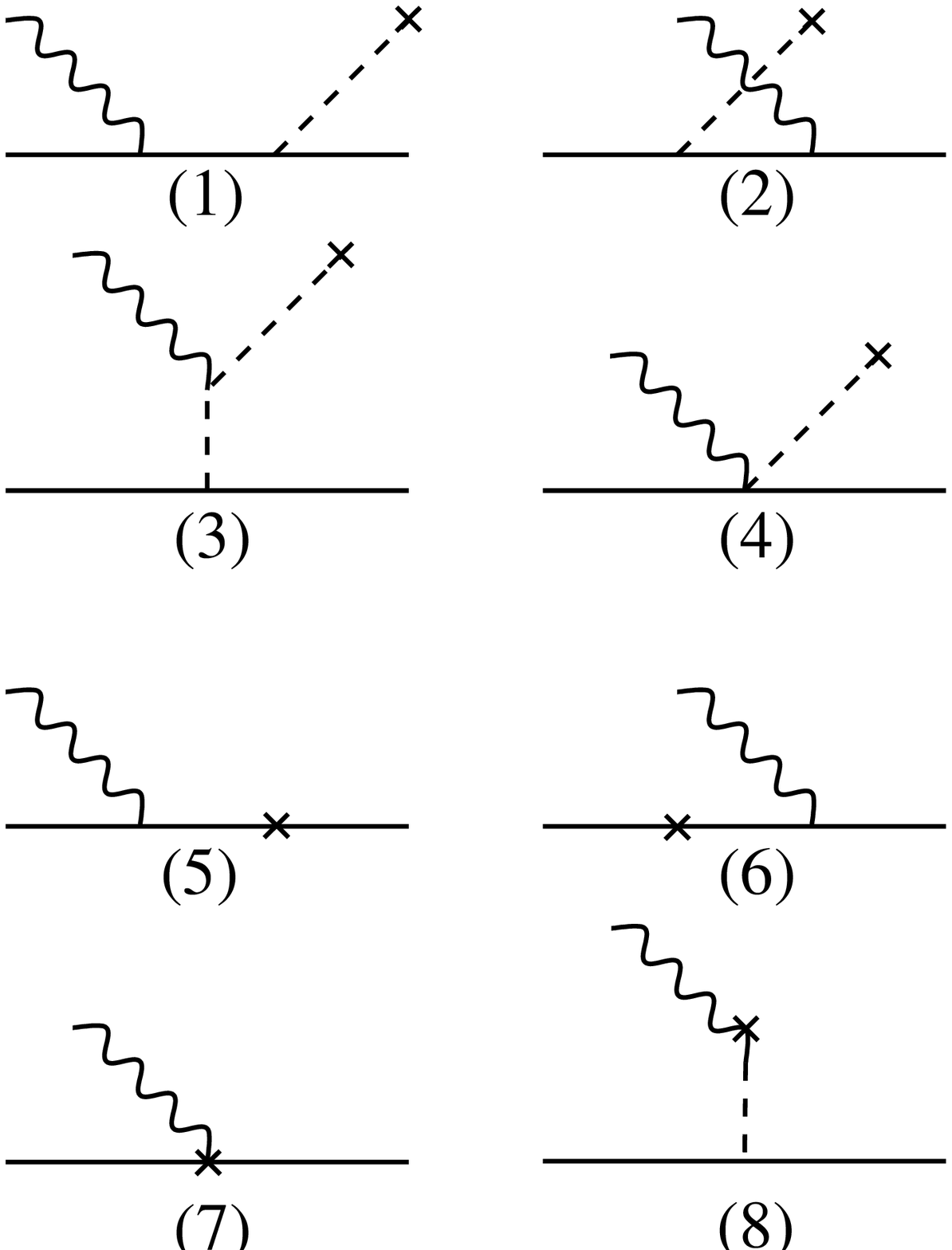, height=10cm}}
   \end{center}
   \caption{\label{mmunujaifig} Feynman diagrams contributing to
the Green function involving the electromagnetic current and
the axial-vector current.
The wavy line denotes a (virtual) photon coupling to the electromagnetic
current. 
   The cross corresponds to the axial-vector current.
   The diagrams have been arranged to yield a maximal similarity with
Fig.\ \ref{mmujpifig}. 
   Note that a diagram of the type (8) is not generated in case of
${\cal M}^\mu_{JP,i}$. 
}
\end{figure}

  Contracting the two $s$-channel diagrams of ${\cal M}^{\mu\nu}_{JA,i}$
with $q_\nu$ [see (1) and (5) of Fig.\ \ref{mmunujaifig}]
we obtain
\begin{eqnarray}
\label{adler1}
q_\nu({\cal M}^{\mu\nu}_{JA,i,1}+{\cal M}^{\mu\nu}_{JA,i,5})&=&
i \left(\frac{F_\pi g_{\pi N}}{m_N} \frac{q^2}{q^2-m_\pi^2}-g_A\right)
\bar{u}(p_f)\left(1-q \dslash \frac{2m_N}{s-m_N^2}
             \right) \gamma_5 \frac{\tau_i}{2}\Gamma^\mu(k) u(p_i) \nn \\ 
          &=& \hat{m}i ({\cal M}^\mu_{JP,i,1}+{\cal M}^\mu_{JP,i,5}).
\end{eqnarray}
   Note that Eq.\ (\ref{adler1}) 
does {\em not} imply a one-to-one correspondence
between diagrams (1) and diagrams (5) of Figs.\ \ref{mmujpifig} and
\ref{mmunujaifig}, respectively.
   In a similar fashion we find for the $u$-channel diagrams
\begin{eqnarray}
\label{adler2}
q_\nu({\cal M}^{\mu\nu}_{JA,i,2}+{\cal M}^{\mu\nu}_{JA,i,6})&=&
i \left(\frac{F_\pi g_{\pi N}}{m_N} \frac{q^2}{q^2-m_\pi^2}-g_A\right)
\bar{u}(p_f)\Gamma^\mu(k)\left(1-q \dslash \frac{2m_N}{u-m_N^2}
             \right) \gamma_5 \frac{\tau_i}{2} u(p_i) \nn \\ 
          &=& \hat{m}i ({\cal M}^\mu_{JP,i,2}+{\cal M}^\mu_{JP,i,6}).
\end{eqnarray}
   Let us now discuss (3) of Fig.\ \ref{mmunujaifig}:
\begin{eqnarray}
\label{adler3}
q_\nu{\cal M}^{\mu\nu}_{JA,i,3}&=&
\epsilon_{3ij}\tau_j \frac{F_\pi q^2}{q^2-m_\pi^2} \frac{(2q-k)^\mu}{t-m_\pi^2}
g_{\pi N} \bar{u}(p_f)\gamma_5 u(p_i)\nonumber\\
&=&\hat{m}i {\cal M}^\mu_{JP,i,3}+\epsilon_{3ij}\tau_j
\frac{(2q-k)^\mu}{t-m_\pi^2}F_\pi
g_{\pi N} \bar{u}(p_f)\gamma_5 u(p_i),\nonumber\\
&\equiv& \hat{m}i {\cal M}^\mu_{JP,i,3}+\Delta {\cal M}^\mu_{i,3},
\end{eqnarray}
   where we introduced the ``remainder'' $\Delta {\cal M}^\mu_{i,3}$
for later purposes.
   In order to obtain Eq.\ (\ref{adler3}) we made use of 
$q^2/(q^2-m_\pi^2)=1+m_\pi^2/(q^2-m_\pi^2)$.
   In a similar way we obtain for (4), (7), and (8) of Fig.\ \ref{mmunujaifig}
\begin{eqnarray}
\label{adler4}
q_\nu{\cal M}^{\mu\nu}_{JA,i,4}&=&
\hat{m}i {\cal M}^\mu_{JP,i,4}+\Delta {\cal M}^\mu_{i,4},\\
\Delta {\cal M}^\mu_{i,4}&=&\epsilon_{3ij}\frac{\tau_j}{2}
F_\pi\bar{u}(p_f)\left\{ \frac{g_{\pi N}}{m_N}\gamma^\mu
+\frac{1}{6}\langle r^2\rangle_A\frac{g_A}{F_\pi}
[(k-q)\cdot k\gamma^\mu -(k-q)^\mu k\hspace{-.5em}/\hspace{.1em}]
\right\} \gamma_5 u(p_i),\\
\label{adler5}
q_\nu{\cal M}^{\mu\nu}_{JA,i,7}&=&
\hat{m}i {\cal M}^\mu_{JP,i,7}+\Delta {\cal M}^\mu_{i,7},\\
\Delta {\cal M}^\mu_{i,7}&=&-\epsilon_{3ij}\frac{\tau_j}{2}
g_A\frac{1}{6}\langle r^2\rangle_A \bar{u}(p_f)[q\hspace{-.5em}/\hspace{.1em}
(q-k)^\mu-\gamma^\mu(q-k)\cdot q]\gamma_5 u(p_i)\nonumber\\
&&-\epsilon_{3ij}\frac{\tau_j}{2}F_\pi \Delta_{\pi N} \frac{g_{\pi N}}{m_N}
\bar{u}(p_f)\gamma^\mu\gamma_5 u(p_i),\\
\label{adler6}
q_\nu{\cal M}^{\mu\nu}_{JA,i,8}&=&\Delta {\cal M}^\mu_{i,8}\nonumber\\
&=&-\epsilon_{3ij}\tau_j g_{\pi N} F_\pi \frac{q^\mu}{t-m_\pi^2}
\bar{u}(p_f)\gamma_5 u(p_i).
\end{eqnarray}
   As a first observation, notice that by construction the sum of 
Eqs.\ (\ref{adler1}) - (\ref{adler6}) adds up to $\hat{m}i {\cal M}^\mu_{JP,i}$
plus the remainders.
   The sum of the latter is given by
\begin{eqnarray}
\Delta{\cal M}^\mu_i&=&
\Delta{\cal M}^\mu_{i,3}+\Delta{\cal M}^\mu_{i,4}
+\Delta{\cal M}^\mu_{i,7}+
\Delta{\cal M}^\mu_{i,8}\nonumber\\
&=&\epsilon_{3ij}\tau_j \frac{(q-k)^\mu}{t-m_\pi^2}F_\pi g_{\pi N}
\bar{u}(p_f)\gamma_5 u(p_i)\nonumber\\
&&+\epsilon_{3ij}\frac{\tau_j}{2} \frac{F_\pi g_{\pi N}}{m_N}
(1-\Delta_{\pi N})\bar{u}(p_f)\gamma^\mu\gamma_5 u(p_i)\nonumber\\
&&+\epsilon_{3ij}\frac{\tau_j}{2}\frac{1}{6}\langle r^2\rangle_A
g_A \bar{u}(p_f)[\gamma^\mu t -(k-q)^\mu (k\hspace{-.5em}/\hspace{.1em}
-q\hspace{-.5em}/\hspace{.1em})]\gamma_5 u(p_i)\nonumber\\
&=&\epsilon_{3ij} \langle N(p_f) | A^\mu_j(0)|N(p_i)\rangle.
\end{eqnarray}
   We thus have explicitly verified the Adler-Gilman
relation, Eq.\ (\ref{gfrel}),
in the framework of the phenomenological approximation to 
${\cal L}_{\rm eff}$.
 
   At this point it is appropriate to recollect that Eq.\ (\ref{gfrel}) is
an {\em exact} relation among QCD Green functions.
   Of course, one would like to verify Eq.\ (\ref{gfrel}) in terms of an ab 
initio QCD calculation.
   On the other hand, a complete and systematic analysis of this chiral Ward 
identity in terms of effective degrees of freedom requires 
effective-field-theory techniques. 
   At this stage one needs the most general effective Lagrangian which is
chirally invariant under {\em local} 
$\mbox{SU}(2)_L\times\mbox{SU(2)}_R\times\mbox{U}(1)_V$ 
transformations provided the external fields are transformed accordingly 
\cite{Gasser:1984yg,Leutwyler:1993iq}.
   This then allows one to deal with the chiral Ward identities in terms of 
an invariance property of the generating functional (see Appendix A of Ref.\ 
\cite{Scherer:2002tk} for a pedagogical illustration). 
   Under these circumstances the chiral Ward identities of QCD (as well as 
their symmetry-breaking pattern) are encoded in the generating functional 
which is then given through the {\em effective} field theory. 
   In the process of constructing the effective Lagrangian one necessarily
also generates non-minimal terms such as, e.g., the $b_{23}$ term in
Eq.\ (\ref{lpin3}) which will be discussed in more detail below.
   As has been illustrated in Ref.\ \cite{Koch:2001ii} 
for the case of the pion electromagnetic vertex, such non-minimal
terms are mandatory for reasons of consistency.

\subsection{Comparison with previous calculations}

\subsubsection{Extrapolation from $q_\mu=0$ to $q_\mu=(m_\pi,0)$}
   The basic idea of traditional current-algebra or PCAC approaches 
consists of defining a function  
\begin{equation}
\label{tildemmui1}
\widetilde{\cal M}^\mu_i(q)
=-i\frac{\hat{m}}{m_\pi^2 F_\pi}(q^2-m_\pi^2){\cal M}^\mu_{JP,i}
\end{equation}
for arbitrary values of $q$,\footnote{Of course, four-momentum conservation
$k+p_i=q+p_f$ is assumed.}
with the property that the physical pion 
production matrix element [see Eq.\ (\ref{limgfrel})] is given by
\begin{equation}
{\cal M}^\mu_i=\left.\widetilde{\cal M}^\mu_i(q)\right|_{q^2=m_\pi^2}.
\end{equation}
   By applying the Adler-Gilman relation, Eq.\ (\ref{tildemmui1}) is 
re-expressed as 
\begin{equation}
\label{tildemmui2}
\widetilde{\cal M}^\mu_i(q)
=\frac{q^2-m_\pi^2}{m_\pi^2 F_\pi}(\epsilon_{3ij} {\cal M}^\mu_{A,j}
-q_\nu{\cal M}^{\mu\nu}_{JA,i}),
\end{equation}
and a constraint for $\widetilde{\cal M}^\mu_i(q)$ is obtained
by evaluating the right-hand side of Eq.\ (\ref{tildemmui2}) at $q_\mu=0$, 
which is traditionally referred to as the soft-pion limit.
   In the present work we prefer the terminology ``soft-momentum limit'' 
which avoids the notion ``off-mass-shell pions.''
   Rather, we consider the Green functions for finite quark masses
(implying massive pions) at $q_\mu=0$, and the result is then translated 
into consistency conditions in terms of the 
invariant amplitudes parameterizing $\widetilde{\cal M}^\mu_i(q)$.

   From the first term of Eq.\ (\ref{tildemmui2}) one obtains the
axial-vector current matrix element for a four-momentum transfer
$k=p_f-p_i$.
   Out of the second term only the one-particle-reducible pole terms are 
candidates contributing to the soft-momentum limit 
\cite{Adler:1965}.\footnote{
We take the soft-momentum limit by first setting $\vec{q}=0$ and then 
performing the limit $q_0 \to 0$.}
   Such $1/q$ singularities in ${\cal M}^{\mu\nu}_{JA,i}$ originate from
pole diagrams where the vertex associated with the axial-current operator
is attached to a nonterminating external nucleon line.
   In principle, these diagrams have to be evaluated using the most 
general renormalized, one-particle-irreducible half-off-shell electromagnetic 
and axial vertices in combination with the most general renormalized
dressed propagator.
   However, expanding vertices and propagators around their on-shell 
values, all such off-shell effects become irrelevant in the soft-momentum
limit.
   This statement requires that none of the off-shell vertices contain poles 
as $q\to 0$. 
   In fact, such poles would have to be of a dynamical origin and are expected
to be absent as long as the underlying dynamics does not contain massless 
particles.\footnote{The prototype of such a pole behavior is, of course, given
by the induced pseudoscalar form factor in case of {\em massless}
pions.}
   Let us illustrate the above statement by use of a ``generic'' axial form 
factor contributing in the $s$-channel diagram,
\begin{eqnarray}
G(q^2,m_N^2,(p_f+q)^2)&=&G(q^2,m_N^2,m_N^2)+(s-m_N^2) G\,'(q^2,m_N^2,m_N^2)
+\cdots,
\end{eqnarray}
with analogous considerations for the electromagnetic vertex. 
   Similarly, the renormalized dressed propagator can be written as
\begin{equation}
S(p_f+q)=S_F(p_f+q)+\mbox{regular terms},
\end{equation}
   where $S_F(p)$ denotes the free propagator of a nucleon with mass $m_N$.
   Finally, noting\footnote{A possible $t$-channel contribution remains
finite, because
\begin{displaymath}
\lim_{q\to 0} \frac{q_\nu}{t-m_\pi^2+i\epsilon}=
   \lim_{q\to 0} \frac{q_\nu}{k^2-m_\pi^2+i\epsilon} = 0.
\end{displaymath}
   Note that in the physical region
$(p_f-p_i)^2\leq 0$ such that the denominator never vanishes. 
}
\begin{mathletters}
\label{qlimits1}
\begin{eqnarray}
\lim_{q\to 0} \frac{q_\nu}{p\hspace{-.5em}/_f+q\hspace{-.5em}/-m_N}
&=&\lim_{q\to 0} 
\frac{q_\nu(p\hspace{-.5em}/_f+q\hspace{-.5em}/+m_N)}{2p_f\cdot q + q^2}
=\frac{g_{\nu 0}(p\hspace{-.5em}/_f+m_N)}{2E_f},\\
\lim_{q\to 0} \frac{q_\nu}{p\hspace{-.5em}/_i-q\hspace{-.5em}/-m_N}
&=&-\frac{g_{\nu 0}(p\hspace{-.5em}/_i+m_N)}{2E_i},
\end{eqnarray}
\end{mathletters}
the soft-momentum limit of $q_\nu {\cal M}^{\mu\nu}_{JA,i}$ reads
\begin{eqnarray}
\label{jmunu_sp}
   \lim_{q \to 0} q_\nu {\cal M}^{\mu\nu}_{JA,i}&=&
   ig_A\bar{u}(p_f)\left[\frac{\gamma^0}{2E_f}\gamma_5 \frac{\tau_i}{2}
   (p\hspace{-.5em}/_f+m_N)  \Gamma^\mu(p_f,p_i)\right.\nonumber\\
   &&\left.\left.
   -\Gamma^\mu(p_f,p_i)(p\hspace{-.5em}/_i+m_N)\frac{\gamma_0}{2E_i}\gamma_5 
   \frac{\tau_i}{2} \right] u(p_i)\right|_{p_f-p_i=k},
\end{eqnarray}
where $\Gamma^\mu(p_f,p_i)$ is given by 
\begin{eqnarray}
\label{gammamuk}
\Gamma^\mu(p_f,p_i)&=&\gamma^\mu \frac{F_1^s(k^2)+\tau_3 F_1^v(k^2)}{2}
+i\frac{\sigma^{\mu\nu}k_\nu}{2m_N}
\frac{F_2^s(k^2)+\tau_3 F_2^v(k^2)}{2},\quad k=p_f-p_i.
\end{eqnarray}
   Here $F^{s/v}_{1/2}(k^2)$ refer to the isoscalar (isovector) Dirac
and Pauli form factors of the nucleon.
   As already pointed out by Adler \cite{Adler:1965}, the positive frequency
projection operators $p\hspace{-.5em}/_f+m_N$ and $p\hspace{-.5em}/_i+m_N$
in the respective $s$- and $u$-channel contributions to Eq.\ (\ref{jmunu_sp}) 
give rise to the fact that only the on-shell electromagnetic vertex enters 
into the soft-momentum result.
   Indeed, we have explicitly checked that inserting the on-shell equivalent
parameterizations involving $G_E$ and $G_M$ or $H_1$ and $H_2$ (for a
discussion see, e.g., Ref.\ \cite{Scherer:1996ux}) generate the 
same soft-momentum limit.
   Moreover, Eq.\ (\ref{jmunu_sp}) only contains the axial-vector
coupling constant $g_A$ but {\em not} the axial form factor. 

   Let us test the consistency of the procedure by contracting
Eq.\ (\ref{jmunu_sp}) with $k_\mu$.
   According to Eq.\ (\ref{cwikmummunujai}) we have
\begin{eqnarray}
k_\mu q_\nu {\cal M}^{\mu\nu}_{JA,i}&=&\epsilon_{3ij}q_\nu
\langle N(p_f)|A^\nu_j(0)|N(p_i)\rangle
=\epsilon_{3ij}[k_\nu-(p_f-p_i)]_\nu 
\langle N(p_f)|A^\nu_j(0)|N(p_i)\rangle
\end{eqnarray}
which clearly vanishes as $q\to 0$, i.e., for $k\to p_f-p_i$.
   On the other hand, from Eq.\ (\ref{jmunu_sp}) we find that both
the $s$- and $u$-channel contributions vanish separately which, of course,
simply reflects the on-shell current conservation condition.

   The consistency relation can be summarized as
\begin{eqnarray}
\label{consrel}
\lefteqn{
\lim_{q\to 0}\widetilde{\cal M}^\mu_i(q)=\left.-\frac{\epsilon_{3ij}}{F_\pi}
\langle N(p_f)|A^\mu_j(0)|N(p_i)\rangle\right|_{p_f-p_i=k}}\nonumber\\
&&\left.
-i\frac{g_A}{F_\pi}\bar{u}(p_f)\left[ \left(1-\frac{m_N}{E_f}\gamma_0\right)
   \gamma_5 \frac{\tau_i}{2} \Gamma^\mu(p_f,p_i)
   +\Gamma^\mu(p_f,p_i) \left(1+ \frac{m_N}{E_i}\gamma_0\right) \gamma_5 
   \frac{\tau_i}{2} \right] u(p_i)\right|_{p_f-p_i=k}.
\end{eqnarray}
  Since the second part of Eq.\ (\ref{consrel}) does {\em not} involve the 
axial form factor, the soft-momentum limit of Eq.\ (\ref{consrel}) leaves
no room for a cancellation of the axial form factor between the axial-vector 
current piece and the second contribution.
   In other words, the soft-momentum limit of $\widetilde{M}^\mu_i(q)$ 
unambiguously contains the axial form factor as well as the induced
pseudoscalar form factor.
   
   Although the specific form of the consistency relation depends on how
the soft-momentum limit is taken, the above conclusion is not affected.
   First of all, the first part of Eq.\ (\ref{consrel}) involving the 
axial-vector current matrix element is path independent,
whereas the directional dependence of the second part is trivial.
   For example, if one wanted to take the soft-momentum limit using
$q^\mu=|\vec{q}\,|(0,\hat{q})$, instead of Eqs.\ (\ref{qlimits1}) 
one would have to consider  
\begin{mathletters}
\label{difflim}
\begin{eqnarray}
\label{difflim1}
\lim_{|\vec{q}\,|\to 0}
&=&\frac{a\cdot q}{p\hspace{-.5em}/_f+q\hspace{-.5em}/-m_N}
=\frac{\vec{a}\cdot\hat{q}}{2\vec{p}_f\cdot\hat{q}}(p\hspace{-.5em}/_f
+m_N),\\
\label{difflim2}
\lim_{|\vec{q}\,|\to 0}
&=&\frac{a\cdot q}{p\hspace{-.5em}/_i-q\hspace{-.5em}/-m_N}
=-\frac{\vec{a}\cdot\hat{q}}{2\vec{p}_i\cdot\hat{q}}(p\hspace{-.5em}/_i
+m_N),
\end{eqnarray}
\end{mathletters}
leading to analogous replacements in the second part of Eq.\ (\ref{consrel}).
   Nevertheless, it would still only contain $g_A$ because the soft-momentum 
limit of functions containing only invariants  has no directional dependence.

   Finally, as an explicit test we evaluate the soft-momentum limit of 
Eq.\ (\ref{tildemmui1}) in the framework of Eqs.\ (\ref{mmujpi}),
\begin{equation}
\lim_{q\to 0}\widetilde{\cal M}^\mu_i(q)=\frac{1}{F_\pi}
\lim_{q\to 0}\, \hat{m}i {\cal M}^\mu_{JP,i},
\end{equation}
   and compare the result with the consistency relation of 
Eq.\ (\ref{consrel}).
   Using
\begin{mathletters}
\label{limits}
\begin{eqnarray}
\label{limit1}
\frac{1}{q^2-m_\pi^2}&\to&-\frac{1}{m_\pi^2},\\
\label{limit2}
1-\frac{2m_N q\hspace{-.5em}/}{s-m_N^2}&\to&1-\frac{m_N}{E_f}\gamma_0,\\
\label{limit3}
1-\frac{2m_N q\hspace{-.5em}/}{u-m_N^2}&\to&1+\frac{m_N}{E_i}\gamma_0,
\end{eqnarray}
\end{mathletters}
together with
$2\hat{m}B=m_\pi^2$ and $(1-\Delta_{\pi N})g_{\pi N}/m_N=g_A/F_\pi$,
   we see that Eq.\ (\ref{mmujpi1}) together with Eq.\ (\ref{mmujpi5}) 
[Eq.\ (\ref{mmujpi2}) together with Eq.\ (\ref{mmujpi6})] exactly
generate the $s$-channel [$u$-channel]  expression of Eq.\ (\ref{consrel}),
whereas the sum of Eqs.\ (\ref{mmujpi3}), (\ref{mmujpi4}), and
(\ref{mmujpi7})
yields
\begin{equation}
\label{matest}
-\frac{1}{F_\pi}\epsilon_{3ij} \bar{u}(p_f)\left[
\gamma^\mu g_A\left(1+\frac{1}{6}\langle r^2\rangle_A k^2\right)
+k^\mu\left(\frac{2 g_{\pi N} F_\pi}{m_\pi^2-k^2}-\frac{m_N g_A}{3}
\langle r^2\rangle_A\right)\right]\gamma_5 \frac{\tau_j}{2}
u(p_i),
\end{equation}
   which corresponds to the nucleon axial-vector current matrix element in 
Eq.\ (\ref{consrel}).
   Thus, the calculation within the framework of ${\cal L}_{\rm eff}$
reproduces the constraint of Eq.\ (\ref{consrel}).

   On the other hand, we would like to emphasize that Eq.\ (\ref{consrel}) 
does not imply a consistency condition for {\em every} pion production
amplitude evaluated for off-shell pion momenta.
   This can easily be visualized by investigating Eqs.\ (\ref{mpepd}) in the 
limit $q\to 0$. 
   We remind the reader that for $q^2\neq m_\pi^2$ the result does not 
correspond to an observable 
\cite{Scherer:1995aq,Fearing:1998wq,Fearing:2000fw} 
but would, for example, be a building 
block of the reaction $\gamma N\to N\gamma\pi$ evaluated in the framework
of ${\cal L}_{\rm eff}$. 
   In fact, the (off-shell) soft-pion limit of ${\cal M}^\mu_i$ looks
similar to $\widetilde{\cal M}^\mu_i(0)$ with the difference
that $g_A/F_\pi$ in the pole terms of Eq.\ (\ref{consrel}) is
replaced by $g_{\pi N}/m_N$.
   The same is true for the $k^2=0$ limit of Eq.\ (\ref{mcontact})
as compared with Eq.\ (\ref{matest}). 
   This is an illustration for the fact that Eq.\ (\ref{consrel}) does not 
yield a consistency relation for the soft-pion production amplitude for an 
{\em arbitrary} interpolating pion field.
   
  At present, corrections to the soft-momentum result of 
Eq.\ (\ref{consrel}) either have to be studied within specific 
models---thus obviously yielding model-dependent results---or can be 
addressed in the framework of ChPT. 
   In the second context such corrections have systematically been 
analyzed at ${\cal O}(p^3)$ in
Refs.\ \cite{Bernard:1991rt,Bernard:1992ys,Bernard:1993rf}, 
where, essentially, a direct calculation of
the pion production matrix element as in Sec.\ V.A was performed.  
   In particular, pion-loop corrections contributing at $q^2=m_\pi^2$ 
modify the soft-momentum result of Eq.\ (\ref{consrel})  
such that the threshold production amplitude $E_{0+}^{(-)}(k^2)$ 
obtains an additional term proportional to 
$m_\pi^2/F_\pi^2$ multiplied by a function $f(k^2/m_\pi^2)$ which vanishes
at $k^2=0$.
   The subtle point about such corrections is that they invalidate 
the naive expectation that corrections to the soft-momentum result
should be of order $m_\pi$ or higher.
   The reason is that pion loops give rise to non-analytic pieces
\cite{Li:1971vr},
where the scale in the loop integrals is set by the pion mass originating 
from the propagators of internal lines.
   Since in ChPT the Green functions are evaluated at a fixed ratio
$\hat{m}/p^2$, the function $f$ counts as ${\cal O}(p^0)$ in the
momentum and quark mass expansion.
   The $m_\pi^2$ in front of the function $f$ reflects
the evolution from the soft-momentum limit $q^2=0$ to $q^2=m_\pi^2$. 

   An {\em explicit} test of the chiral Ward identity of Eq.\ (\ref{gfrel}) at 
${\cal O}(p^3)$ including the loop corrections is not yet available in the 
literature.

\subsubsection{Comparison with Haberzettl}

   Recently, the question whether the axial radius of the nucleon
can be obtained from threshold pion electroproduction data 
\cite{Choi:1993vt,Blomqvist:tx,Liesenfeld:1999mv,Bartsch:1998,Mueller:2002,%
Baumann:2003} has given rise 
to much controversy \cite{Haberzettl:2000sm,Guichon:2001ew,Bernard:2001ii,%
Haberzettl:2001yt,Truhlik:2001nn}.
   The discussion was triggered by a paper of Haberzettl 
\cite{Haberzettl:2000sm}, where it was argued that PCAC does not
provide any additional constraints beyond the Goldberger-Treiman relation.
   Similar claims were made by Ohta in Ref.\ \cite{Ohta:1992ui} some time 
ago.

   In order to solve this seeming puzzle we need to have a closer look
at the method used in Ref.\ \cite{Haberzettl:2000sm}.
   Starting from the nucleon matrix element of the axial-vector current
[see Eq.\ (\ref{axial_current})] in combination with the constraint
of Eq.\ (\ref{ff_relation}), the axial-vector current was split into 
``weak'' and ``hadronic'' parts, expressed in terms of $G_A$ and 
$G_{\pi N}$, respectively.
   Such a splitting may be interpreted as resulting from
the (formal) separation of the axial-vector current operator into a
transversal part and a longitudinal one \cite{Furlan:1966},
\begin{equation}
\label{axsep}
A^\mu_i(x)=\left[A^\mu_i(x)-\frac{\partial_\mu\partial_\nu}{\Box} A^\nu_i(x)
\right]+\frac{\partial_\mu\partial_\nu}{\Box} A^\nu_i(x).
\end{equation}
   After this separation a formal expression for (the equivalent of) the 
Green function ${\cal M}_{JA, i}^{\mu\nu}$ of Eq.\ (\ref{g2a}) was constructed.
   This was done by inserting an external photon in all possible places
in the diagram corresponding to the separation of the axial-vector
current (see Figs.~1 and 3 of Ref.\ \cite{Haberzettl:2000sm}).
   For the insertion {\em into} vertices the so-called gauge-derivative
method of Ref.\ \cite{Haberzettl:1997jg} was applied.
   For example, for the last diagram of Fig.~3 of 
Ref.\ \cite{Haberzettl:2000sm} corresponding to diagram 4 of our 
Fig.\ \ref{mmunujaifig} one needs the contact interaction of pion 
electroproduction as obtained
from the insertion into the pion-nucleon vertex.
   For our case, this vertex is given by Eq.\ (\ref{pnvertex}), and the
application of the gauge-derivative method would simply produce the contact 
vertex
\begin{equation}
\label{contactvertex}
ie\epsilon_{3ij}\tau_j \frac{g_{\pi N}}{2m_N}\gamma^\mu\gamma_5.
\end{equation}
   Of course, in the present case, Eq.\ (\ref{contactvertex}) is nothing else 
than what is generated by minimal substitution into the pseudovector pion
nucleon interaction.
   However, this is {\em not} what chiral symmetry tells us. 
   In order to see this we have to compare with the result for the 
$\gamma \pi NN$ vertex of Eq.\ (\ref{mcontact}), namely
\begin{equation}
    ie\epsilon_{3ij}\tau_j\left\{ \frac{g_{\pi N}}{2m_N}\gamma^\mu +
    \frac{g_A}{12F_\pi} \ra
    \left[ (k-q)\cdot k \, \gamma^\mu - (k-q)^\mu k \dslash \right] \right\}
    \gamma_5.
\end{equation}
   We conclude that the gauge derivative-method produces only part of the
full interaction and is in conflict with the constraints of chiral symmetry.
   In the above case it does not generate the $\ra$ term entering the
charged-pion electroproduction amplitude.

\subsection{The role of chiral symmetry}
   From the effective-field-theory point of view it is rather straightforward
to understand how a quantity such as $\ra$ enters different physical 
amplitudes.
   Due to spontaneous symmetry breaking, the chiral symmetry of QCD
is realized nonlinearly on the effective degrees of freedom 
\cite{Weinberg:1979kz,Gasser:1984yg,Gasser:1984gg,Gasser:1988rb,%
Weinberg:1968de,Coleman:sm} [see Eqs.\ (\ref{utrans}) and (\ref{ntrans})].
   In order to collect the chiral Ward identities in a generating functional
one needs the most general locally invariant effective 
Lagrangian where the emphasis is on both {\em generality} and {\em
local invariance}.
   In the present case we will have a closer look at the $b_{23}$ term 
of Eq.\ (\ref{lpin3}) involving the quantity $f_{\mu\nu}^-$ of Eq.\
(\ref{fpm}): 
\begin{eqnarray}
  f_{\mu\nu}^- &=& u\left\{ \partial_\mu (v_\nu-a_\nu) - 
     \partial_\nu (v_\mu-a_\mu) -i[v_\mu-a_\mu,v_\nu-a_\nu] \right\} u^\dagger
\nn \\ && - u^\dagger \left\{ \partial_\mu (v_\nu+a_\nu) - 
     \partial_\nu (v_\mu+a_\mu) -i[v_\mu+a_\mu,v_\nu+a_\nu] \right\} u \nn\\
   &=& -2(\partial_\mu a_\nu - \partial_\nu a_\mu) + 2i \left(
       [v_\mu, a_\nu] - [v_\nu, a_\mu] \right) \nn\\
   && +\frac{i}{F} \left[\vec{\tau}\cdot\vec{\pi}, 
      \partial_\mu v_\nu - \partial_\nu v_\mu
   \right]
      + \frac{1}{F} \left[\vec{\tau}\cdot\vec{\pi}, [v_\mu,v_\nu] \right]
      + \frac{1}{F} \left[\vec{\tau}\cdot\vec{\pi}, [a_\mu,a_\nu] \right] 
      + O(\pi^2),
\end{eqnarray}
where we expanded $u$ in terms of the pion field.
   We first note that $f_{\mu\nu}^-$ involves field-strength tensors as 
opposed to pure covariant-derivative terms.
   Moreover, due to the nonlinear realization it contains a string of
terms with an increasing number of pion fields.    
   The lowest-order term involving one external axial-vector field,
$$-2(\partial_\mu a_\nu - \partial_\nu a_\mu),$$
gives rise to a contribution to the axial-vector current matrix element.
   It is responsible for the identification of the $b_{23}$ term with the
axial radius.
   On the other hand, there is no term with only one pion, i.e., no 
contribution to the $\pi NN$ vertex of Eq.\ (\ref{pnvertex}). 
   In addition, there is also no contribution to the strong form
factor $G_{\pi N}$ of Eq.\ (\ref{gt1}).
   The term
$$\frac{i}{F} \left[\vec{\tau}\cdot\vec{\pi}, 
\partial_\mu v_\nu - \partial_\nu v_\mu\right]$$
contributes to the $\gamma \pi NN$ vertex. 
   Thus, we clearly see how chiral symmetry relates for this particular
term (a part of) the axial-vector current vertex with (a part of) the 
$\gamma \pi NN$ vertex.
   On the other hand, this relation is not generated by the
gauge-derivative method.

\section{Summary and conclusions}
\label{section_conclusions}
   We have re-investigated Adler's PCAC relation in the presence of an
external electromagnetic field \cite{Adler:1965} within the framework of 
QCD coupled to external fields \cite{Gasser:1984yg,Gasser:1984gg}.
   With a suitable choice for the interpolating pion field the QCD result 
is of the same form as Adler's pre-QCD version.
   We then discussed the Adler-Gilman relation \cite{Adler:1966} as a 
chiral Ward identity in terms of QCD Green functions and established
the connection with the pion electroproduction amplitude.
   In order to explain the consequences of the Adler-Gilman relation, we made
use of a tree-level approximation to the Green functions at ${\cal O}(p^3)$
within relativistic baryon chiral perturbation theory. 
   As a reference point we first performed a direct calculation of the 
pion-production transition current, ${\cal M}^\mu_i$, in terms of the 
effective degrees of freedom.
   We saw explicitly how the axial radius enters charged-pion
electroproduction at ${\cal O}(p^3)$.
   As an alternative we calculated the Green function 
${\cal M}_{JP, i}^\mu$ involving the 
electromagnetic current and the pseudoscalar density and, using
the LSZ reduction formalism, explicitly verified the connection with the pion 
electroproduction transition current determined previously.
   Again we saw that the axial radius enters this particular Green function.
   As a test of our result we verified a chiral Ward identity relating
the divergence of ${\cal M}_{JP, i}^\mu$ to the matrix element of the
pseudoscalar density.
   We then calculated the Green function ${\cal M}_{JA, i}^{\mu\nu}$ involving
the electromagnetic and axial-vector currents, tested the constraints due
to gauge invariance, and, finally, explicitly verified the Adler-Gilman
relation for {\em arbitrary} values of $q^2$.
   Thus, all three possibilities of calculating pion electroproduction---%
direct calculation, determination in terms of the QCD Green function
${\cal M}_{JP, i}^\mu$, or application of the Adler-Gilman relation---%
generate the same result.

   We then made contact with the traditional current-algebra or PCAC techniques
by defining a generalization $\widetilde{\cal M}^\mu_i$ of the physical
pion electroproduction transition current in terms of the
QCD Green function ${\cal M}_{JP, i}^\mu$ for arbitrary values of $q$.
   We considered the soft-momentum limit of $\widetilde{\cal M}^\mu_i$, 
$q^\mu\to 0$, and showed that the usual ``soft-pion'' results 
are recovered if the pseudoscalar density is used as the pion interpolating 
field. 
   We pointed out how the nonlinear realization of chiral symmetry leads
to an interplay between various vertices in the most general theory
and how approximations such as minimal substitution may fail to be
compatible with the strictures of chiral symmetry and can lead to 
erroneous conclusions.
   
   Clearly, chiral perturbation theory has become the standard method to 
systematically deal with {\em corrections} to the current-algebra results
beyond the phenomenological approximation.
   The contribution of loops diagrams are expected to separately satisfy the 
constraints due to the Ward identities.
   In the case of pion photo- and electroproduction such corrections 
were determined in Refs.\ 
\cite{Bernard:1991rt,Bernard:1992ys,Bernard:1993rf,Bernard:2001rs}
leading to additional terms beyond the 
current-algebra results.
   Obviously, it would be nice to have a fully relativistic calculation 
within the infrared regularization \cite{Becher:1999he} 
or the extended-on-mass-shell scheme \cite{Fuchs:2003qc}
including an explicit test in terms of the Adler-Gilman relation.

\acknowledgements 
  The authors would like to thank J.~Gegelia and J.~H.~Koch for useful
discussions. 
  This work was supported by the Deutsche Forschungsgemeinschaft (SFB 443).

\end{document}